\documentclass[12pt,preprint]{aastex}

\shorttitle{Testing the CMB Data for Systematic Effects}
\shortauthors{Griffiths \& Lineweaver}

\received{2003 July 28}
\begin{document}

\title{Testing the CMB Data for Systematic Effects}

\author{Louise M. Griffiths and Charles H. Lineweaver}

\affil{Department of Astrophysics and Optics, School of Physics, University of New South Wales, Sydney, NSW 2052, Australia}

\begin{abstract}
Under the assumption that the concordance $\Lambda$ cold dark matter
(CDM) model is the correct model, we test the cosmic microwave
background (CMB) anisotropy data for systematic effects by examining
the band pass temperature residuals with respect to this model.
Residuals are analysed as a function of angular scale $\ell$, galactic
latitude, frequency, calibration source, instrument type and several
other variables that may be associated with potential systematic
effects. Our main result is that we find no significant systematic 
errors associated with these variables.  
However, we do find marginal evidence for a trend associated with
galactic latitude indicative of galactic contamination.
\end{abstract}

\keywords{cosmic microwave background - cosmology: observations}

\section{INTRODUCTION}
The cosmic microwave background (CMB) power spectrum is a particularly
potent probe of cosmology.  As long as the systematic errors
associated with these observations are small, the detected signal has
direct cosmological importance. The ever-tightening network of
constraints from CMB and non-CMB observations favours a concordant
$\Lambda$ cold dark matter (CDM) model that is commonly accepted as
the standard cosmological model (Table~\ref{concord}).  Since the
anisotropy power spectrum is playing an increasingly large role in
establishing and refining this model, it is crucial to check the CMB
data for possible systematic errors in as many ways as possible.

Systematic errors and selection effects are notoriously difficult to
identify and quantify.  Individual experimental groups have developed
various ways to check their CMB observations for systematic effects
\citep[e.g.][]{kogut96,miller02}, including the use of multiple
calibration sources, multiple frequency channels and extensive beam
calibrating observations. Internal consistency is the primary concern
of these checks.

Testing for consistency with other CMB observations is another
important way to identify possible systematic errors.  When the areas
of the sky observed overlap, this can be done by comparing CMB
temperature maps \citep[e.g.][]{ganga94a,lineweaveretal95,xu01}. When
similar angular scales are being observed one can compare power
spectra \citep[e.g.][Figure 11]{sievers02}.  A prerequisite for the
extraction of useful estimates for cosmological parameters from the
combined CMB data set is the mutual consistency of the observational
data points \citep{wangetal02a}; the best-fit must also be a good
fit. \citet{wangetal02a} and \citet{sievers02} have recently explored
the consistency of various CMB observations with respect to power
spectrum models and concluded that the CMB fluctuation data is
consistent with several minor exceptions.

Although individual observational groups vigorously test their data
sets for systematic errors, the entire CMB observational data set has
not yet been collectively tested.  Here we check for consistency of
the concordance model (Table~\ref{concord}) with respect to possible
sources of systematic error.  Under the assumption that the
concordance model is the correct model (i.e. more correct than the
best-fit to the CMB data alone), we explore residuals of the
observational data with respect to this model to see if any patterns
or outliers emerge.  We attempt to identify systematic errors in the
data that may have been ignored or only partially corrected for.

With only a few independent band power measurements the usefulness of
such a strategy is compromised by low number statistics. However, we
now have hundreds of band power measurements on scales of
$2<\ell<2000$ from over two dozen autonomous and semi-autonomous
groups. There are enough CMB fluctuation detections from independent
observations that subtle systematic effects could appear above the
noise in regression plots of the data residuals.  This is particularly
the case when one has a better idea of the underlying model than
provided by the CMB data alone.

The history of the estimates of the position of the CMB dipole
illustrates the idea.  Once a relatively precise direction of the
dipole was established, the positional scatter elongated in the
direction of the galactic centre could be distinguished unambiguously
from statistical scatter and more reliable corrections for galactic
contamination could be made \citep[Figure 2]{lineweaver97}.  We aim to
ascertain whether the use of the concordance model as a prior can help
to separate statistical and systematic errors in the CMB anisotropy
data.  In \S2 we discuss constraints on cosmological parameters, the
current concordance model and how simultaneously analysing
combinations of independent observational data sets can tighten
cosmological constraints. Our analytical methodology is detailed in
\S3.  In \S4 possible sources of systematic uncertainty are discussed.
In \S5 and \S6 our results are discussed and summarised.

\section{THE CONCORDANCE COSMOLOGY}
\subsection{Observational concordance} 
The CMB has the potential to simultaneously constrain a number of
cosmological parameters that are the ingredients of the hot big bang
model.  Unfortunately, particular parameter combinations can produce
indistinguishable $C_{\ell}$ spectra \citep{efbond99}.  For example,
cosmological models with different matter content but the same
geometry can have nearly identical power spectra.  Such model
degeneracies limit parameter extraction from the CMB alone.

A number of recent analyses combine information from a range of
independent observational data sets
\citep[e.g.][]{efstetal02,lewisbridle02,sievers02,wangetal02b,spergel03},
enabling certain degeneracies of the individual data sets to be
resolved.  As the observational data become more precise and diverse
they form an increasingly tight network of parameter constraints.
Constraints from a variety of astrophysical data, including CMB
temperature (TT) and temperature-polarisation (TE) angular power
spectra, the 2-degree Field Galaxy Redshift Survey (2dFGRS) power
spectrum \citep{percival01}, supernova type Ia (SNIa) measurements of
the angular diameter distance relation \citep{garnavich98,riess01},
measurements of the Lyman alpha (Ly$\alpha$) forest power spectrum
\citep{croft02,gnedham02}, Hubble Key Project (HKP) constraints on the
Hubble parameter $h$ \citep{freedman01} and Big Bang nucelosynthesis
(BBN) constraints on the baryon fraction \citep{burlesetal01}, are
beginning to refine an observationally concordant cosmological model.

The results of recent CMB-only analyses and joint likelihood analyses
are given in Table~\ref{concord}.  The most current joint analysis to
date suggests the observationally concordant cosmology
\citep{spergel03}; $\Omega_{\kappa}\simeq0$, $\Omega_{\Lambda}\simeq
0.7$ ($\Omega_{m} = \Omega_b+\Omega_c\simeq 0.3$), $\Omega_b h^2
\simeq0.0226$, $n_s \simeq 0.96$, $h \simeq 0.72$ and $\tau \simeq
0.12$ with $A_t$, and $\Omega_{\nu}$ taken to be zero.  With more
precise and diverse cosmological observations, the ability of this
standard $\Lambda$CDM cosmology to describe the observational universe
will be extended and tested for inconsistencies.

\subsection{Goodness of fit of the concordance model to the CMB}
We perform a simple $\chi^{2}$ calculation (see Appendix~A) to
determine the goodness-of-fit of the concordance $\Lambda$CDM
cosmology to the CMB, employing the band power temperature
measurements of Table~\ref{obsdat} and their associated window
functions.  We limit our analysis to $2<\ell<2000$ because secondary
anisotropy contributions, such as the Sunyaev-Zel'dovich effect
\citep{sunzel70} and/or the signature of primordial voids
\citep{griffithsetal03}, may dominate at $\ell>2000$.  We bin the WMAP
data into 70 bins, carefully chosen so as not to smooth out any
genuine features in the data, following the method of Appendix~B.

The model radiation angular power spectrum is calculated using {\sc
cmbfast} \citep{selzal96}.  However, rather than adopting the {\sc
cmbfast} COBE-DMR normalisation, we implement a numerical
approximation to marginalisation (see Appendix A) to find the optimal
normalisation of the theoretical model to the full observational data
set.  We also similarly treat the beam uncertainties of BOOMERanG98,
MAXIMA1 and PyV and the calibration uncertainties associated with the
observations as free parameters with Gaussian distributions (see
Eq.~\ref{e:chisq}).

The minimised $\chi^2$ for the concordance model is 258. In order to
determine how good a fit this model is to the observational data we
need to know the number of degrees of freedom of the
analysis. Although 274 degrees of freedom are provided by the number
of observational data points (assuming they are uncorrelated), these
are reduced by the number of concordance parameters that are
constrained using the CMB data alone.  The flatness of the concordance
model ($\Omega_{\kappa} \simeq 0$), the tilt of the primordial power
spectrum of scalar perturbations ($n_s \simeq 0.96$) and the optical
depth of reionisation ($\tau \simeq 0.12$) are extracted almost
entirely from the CMB data.  The remaining concordant parameters are
strongly constrained by non-CMB observations.  We therefore estimate
that 3 degrees of freedom should be subtracted from the original 274.

Within our analysis we marginalise over a number of nuisance
parameters.  We fit for 22 individual calibration constants (all the
CBI observations are assumed to shift together), 3 beam uncertainties
(those of BOOMERanG-98, MAXIMA-1 and PyV) and an overall
normalisation.  Thus a further 26 degrees of freedom must be
subtracted leaving 245 degrees of freedom.  The $\chi^2$ per degree of
freedom is then $258/245=1.05$, indicating that the concordance cosmology
provides a good fit to the CMB data alone.

Data correlations other than the correlated beam and calibration
uncertainties of individual experiments, that we take to have no
inter-experiment dependence, are not considered in our analysis.
Including such correlations would further reduce the number of degrees
of freedom, increasing the $\chi^2$ per degree of freedom.  However,
our result is in agreement with joint likelihood analyses that
find that the cosmological model that best fits the CMB data is a
better fit at the 1- or 2-$\sigma$ level than the fit to the concordance
model \citep{wangetal02a}.

We find the normalisation of the concordance model to the full CMB
data set to be $Q_{10} = 18.56 \pm 0.04 \,\mu$K, where $Q_{10}$ is
defined through the relation \citep{linbar98},
\begin{equation}
10(10+1)C_{10} = \frac{24\pi}{5} \frac{Q_{10}^2}{T_{CMB}^2} \,.
\end{equation}
The normalised concordance model is plotted with the calibrated and
beam corrected observational data in Figure~\ref{clplotcol}. It is
difficult to distinguish the most important measurements because there
are so many CMB data points on the plot and it is dominated by those
with the largest error bars.  Therefore, for clarity, we bin the data
as described in Appendix~B.  The binned observations are plotted with
the concordance cosmology in Figure~\ref{bin_clplotcol}. 

\section{EXAMINING THE RESIDUALS}
Our analysis is based on the assumption that the combined cosmological
observations used to determine the concordance model are giving us a
more accurate estimate of cosmological parameters, and therefore of
the true $C_{\ell}$ spectrum, than is given by the CMB data
alone. Under this assumption, the residuals of the individual observed
CMB band powers and the concordance $\Lambda$CDM model become tools to
identify a variety of systematic errors.  To this end, we create
residuals $R_i$ of the observed band power anisotropies
$C_{\ell_{\rm eff}}^{\rm obs}(i)\pm\sigma^{\rm obs}(i)$ with respect to
the concordant band powers $C_{\ell_{\rm eff}}^{\rm th}(i)$ such that,
\begin{equation}
\label{e:resid}
R_i = \frac{C_{\ell_{\rm eff}}^{\rm obs}(i)-C_{\ell_{\rm eff}}^{\rm
th}(i)}{C_{\ell_{\rm eff}}^{\rm th}(i)} \pm \frac{\sigma^{\rm
obs}(i)}{C_{\ell_{\rm eff}}^{\rm th}(i)} \,.
\end{equation}

Systematic errors are part of the CMB band power estimates at some
level.  We examine our data residuals as functions of the instrumental
and observational details, listed in Table~\ref{obstech}, that may be
associated with systematic errors. Possible sources of systematic
uncertainty are discussed in the following section.  If the analysis
determines that a linear trend can produce a significantly improved
fit in comparison to that of a zero gradient line (zero-line) through
the data, it may be indicative of an unidentified systematic source of
uncertainty.  Similarly, any significant outliers may point to
untreated systematics. Our results are summarised in
Table~\ref{resres}.

The zero-line through all the residual data gives a $\chi^2$ of 258.
As previously discussed, the analysis that determines the
goodness-of-fit of the concordance model to the CMB data has 245
degrees of freedom. A further 2 degrees of freedom must be subtracted
for the line slope and intercept parameters that are varied in the
line fitting analysis, leaving 243 degrees of freedom.  However, when
the residuals are examined as a function of angular scale, the
intercept of any line that fits the residual data will depend on the
normalisation of the concordance model. In this case, we therefore
subtract only one further degree of freedom, giving 244 degrees of
freedom.

The zero-line fit to the residual data has a $\chi^2$ per degree of
freedom of 1.06 and a 76\% probability of finding a line that better
fits the data.  In order to determine the significance of a better fit
provided by a linear trend, an understanding of the statistical
effects of introducing the 2 parameters to the line-fitting analysis
is required.  For a 2-dimensional Gaussian distribution, the
difference between the $\chi^2$ of the best-fit model and a model
within the 68\% confidence region of the best-fit model is less than
2.3 and for a model that is within the 95\% confidence region of the
best-fit model, this difference is less than 6.17 \citep{numrec}.  Our
68\% and 95\% contours in Figures~\ref{lresid} to \ref{pointresid} are
so defined.  The further the horizontal concordance zero-line is from
the best-fitting slope, the stronger the indication of a possible
systematic error.

\section{POSSIBLE SOURCES OF SYSTEMATIC UNCERTAINTY}
\subsection{Angular scale-dependent effects}
We examine scale-dependent uncertainties by plotting the residuals as
a function of $\ell$ (Figure \ref{lresid}).  The shape of the window
function is most critical when the curvature of the power spectrum is
large (at the extrema of the acoustic oscillations). We therefore also
explore the residuals as a function of the narrowness of the filter
functions $\Delta \ell/\ell$.

The resolution of the instrument and the pointing uncertainty become
increasingly important as fluctuations are measured at smaller angular
scales.  Small beams may be subject to unidentified smearing effects
that may show up as a trend in the residual data with respect to
$\theta_{\rm beam} \ell_{\rm eff}$.  Thus we examine the residuals as a
function of $\theta_{\rm beam} \ell_{\rm eff}$ (Figure \ref{resresid}) and
pointing uncertainty (Figure \ref{pointresid}) to look for hints of
systematic errors associated with these factors.

\subsection{Foregrounds}
If foreground emission is present, it will raise the observed power.
Galactic and extra-galactic signals from synchrotron, bremsstrahlung
and dust emission have frequency dependencies that are different from
that of the CMB \citep[e.g.][]{tegef96}. If such contamination is
present in the data, it may be revealed by a frequency dependence of
the residuals (Figure \ref{nuresid}).

Multiple frequency observations provide various frequency lever-arms
that allow individual groups to identify and correct for frequency
dependent contamination.  Experiments with broad frequency coverage
may be better able to remove this contamination than those with narrow
frequency coverage.  We therefore examine the residuals as a function
of the frequency lever-arm $(\nu_{\rm max}-\nu_{\rm min})/\nu_{\rm
main}$ (Figure \ref{deltanuresid}).

Observations taken at lower absolute galactic latitudes, $|b|$, will
be more prone to galactic contamination. We check for this effect by
examining the residuals as a function of $|b|$ range
(Figure~\ref{gallatresid}).  In this case, we take $|b|$ to be the
value central to the range observed and the 1-$\sigma$ uncertainties
in $|b|$ to extend to the extrema of this range (see
Table~\ref{obstech}). We also examine the residuals as a function of
central $|b|$ neglecting the range in $|b|$ observed
(Figure~\ref{gallatresidb}). Using ranges in $|b|$ as statistical
errors as in Figure~\ref{gallatresid} is problematic, but so too is
treating the band powers as if they result from measurements at a
precise value of $|b|$ (Figure~\ref{gallatresidb}).  The most plausible result is
intermediate between these two cases.

\subsection{Calibration}
To analyse various experiments, knowledge of the calibration
uncertainty of the measurements is necessary.  Independent
observations that calibrate off the same source will have calibration
uncertainties that are correlated at some level and therefore a
fraction of their freedom to shift upwards or downwards will be
shared.  For example, ACME-MAX, BOOMERanG97, CBI, MSAM, OVRO, TOCO and
CBI all calibrate off Jupiter, so part of the quoted calibration
uncertainties from these experiments will come from the brightness
uncertainty of this source. \citet{wangetal02a} perform a joint
analysis of the CMB data making the approximation that the entire
contribution to the calibration uncertainty from Jupiter's brightness
uncertainty is shared by the experiments that use this calibration
source.  The true correlation will be lower since the independent
experiments observed Jupiter at different frequencies.

Inter-experiment correlations are not considered in our analysis,
since we are unable to separate out the fraction of uncertainty that
is shared by experiments.  Instead, we test for any calibration
dependent systematics by examining the data residuals with respect to
the calibration source (Figure \ref{csresid}). We note that including
correlations between data points would reduce the number of degrees of
freedom of our $\chi^2$ analysis.  The order of the calibration
sources is arbitrary so the fitting of a line serves only to verify
that the line-fitting and confidence-interval-determining codes are
working as expected.  However, any significant outliers may indicate
unidentified calibration dependent systematics.

\subsection{Instrument type and platform}
The experiments use combinations of 3 types of detector that operate
over different frequency ranges.  We classify the data with respect to
their instrument type; HEMT interferometers (HEMT/Int), HEMT amplifier
based non-interferometric instruments (HEMT), HEMT based amplifier and
SIS based mixer combination instruments (HEMT/SIS), bolometric
instruments and bolometric interferometers (Bol/Int). We check for
receiver specific systematic effects by plotting the residuals as a
function of instrument type (Figure \ref{typeresid}). Again the order
we choose for instrument type is arbitrary and it is significant
outliers that we are interested in.

Water vapour in the atmosphere is a large source of contamination for
ground based instruments.  There may also be systematic errors
associated with the temperature and stability of the thermal
environment.  We therefore explore instrument platform dependencies of
the data residuals.  We choose to order the instrument platform
according to altitude.

\subsection{Random controls}
We use a number of control regressions to check that our analysis is
working as expected.  To this end, the residuals are examined with
respect to the publication date of the band power data, the number of
letters in the first author's surname and the affiliation of the last
author.  We expect the line fitted to these control regressions to be
consistent with a zero-line through the residual data.  Any
significant improvement provided by a linear fit to these residuals
may be indicative of a problem in the software or methodology.

\section{RESULTS}
For the regressions plotted, the residual data is binned as described
in Appendix~B so that any trends can be more effectively visualised.
Since the data binning process may wash out any discrepancies between
experiments, the linear fit analyses are performed on the unbinned
data residuals.  In Figures~\ref{lresid} to \ref{pointresid}, the line
that best-fits the data is plotted (solid white) and the 68\% (dark
grey) and 95\% (light grey) confidence regions of the best-fit line
are shaded.  For the plots for which it makes sense to test for a
linear dependence, we report the $\chi^2$ per degree of freedom for
the best-fit line and comment on the significance of the deviation of
the zero-line (dashed black).  For those plots for which the $x$-axis
order is arbitrary, we comment on any significant outliers from the
best-fit line.

Our results are summarised in Table~\ref{resres}.  The lines fitted to
our control regressions are consistent with a zero-line through the
residual data suggesting that our line-fitting and
confidence-interval-determining codes are working as expected.  We
find a linear trend in the residuals with respect to the
$|b|$ range of the observations (see Figure~\ref{gallatresid}).  This
trend is not eliminated by the removal of any one experiment and may
be indicative of a source of galactic emission that has not been
appropriately treated. 

In Figure~\ref{gallatresid}, the errors in both the $y$- and
$x$-directions are used in the fit. We have defined $|b|$ to be that
of the centre of the observations and the uncertainties to extend to
the edges of the range.  This allows the observations some freedom of the
$x$-coordinate in the line-fitting analysis and weights heavily those
detections that span small ranges in absolute galactic latitude. It is
therefore also interesting to examine the residuals with respect to
the central $|b|$ to determine the significance of the trend with the
$x$-coordinate freedom removed (see Figure~\ref{gallatresidb}). The
most plausible galactic latitude regression will be somewhere between
the regressions shown in these two plots.

Removing the $x$-coordinate freedom removes the significance of the
trend. This result implies that experiments that observe over small
ranges in galactic latitude are dominating the trend and we therefore
can not simply correct for the systematic that is implied in
Figure~\ref{gallatresid}.  The comparison of rms levels in galactic
dust \citep{finkbeiner99} and
synchrotron\footnote{http://astro.berkeley.edu/dust} maps over the
areas of CMB observations may help to clarify the interpretation of
the trend.  Such a technique has recently been applied to the MAXIMA1
data \citep{jaffeetal03} but has yet to be performed on the full CMB
data set.

Other plots also show some evidence for systematic
errors. Figures~\ref{bin_clplotcol} and \ref{lresid} indicate that 6
bins at $\ell > 900$ prefer a lower normalisation. This may be due to
a systematic calibration error for some of the experiments in this
$\ell$ range, underestimates of beam sizes or pointing uncertainties
or unidentified beam smearing effects at high $\ell$ for small beams.
Although, Figures \ref{csresid} and \ref{pointresid} show little
evidence for any trends, Figure \ref{resresid} shows marginal evidence
for power suppression at low $\theta_{\rm beam}\ell_{\rm eff}$.  The
motivation for this plot is to see if there are any systematics
associated with large beams sampling small scale anisotropies (right
side of plot) or with small beams sampling large scale anisotropies
(left side of plot). Small beams used to measure large angular scales
may have stability problems analogous to the problems one runs into
when trying to mosaic images together.  Although no overall linear
trend is observed, the trend indicated for $\theta_{\rm beam}\ell_{\rm
eff} < 900$ suggests that increased attention should be given to band
power estimates in this regime.

\section{SUMMARY}
Over the past 10 years, successive independent and semi-independent
data sets have extended the angular scale, calibration precision and
freedom from galactic contamination of the CMB power spectrum. Each
CMB measurement contains useful cosmological information and no data
set is immune to contamination. It is therefore important to compare
data sets and check for systematics. We have collectively tested the
full CMB data set for inconsistencies with the concordance model and
our results indicate that the model is consistent with the data
although a need to slightly dampen power in the model at high $\ell$
is indicated (Figure~\ref{lresid}).

We have explored residuals of the observational data with respect to
the concordance model to see if any patterns emerge that may indicate
a source of systematic error. We have found little significant
evidence for inter-experiment inconsistencies other than 
a trend associated with galactic latitude that may be an indication
of low-level galactic contamination of CMB observations made 
closer to the galactic plane (Figure~\ref{gallatresid}).  
A more detailed comparison of CMB fields of observations with
galactic dust and synchrotron maps will be necessary to
clarify the source of this trend.

\acknowledgments LMG thanks Martin Kunz for useful discussions and is
grateful to the University of Sussex where part of the work was
carried out.  LMG acknowledges support from the Royal Society and
PPARC. CHL acknowledges a research fellowship from the Australian
Research Council.

\clearpage

\appendix

\section{$\chi^2$ MINIMISATION METHODOLOGY}
For the observational power spectrum data quoted in the literature,
individual $C_{\ell \,}$s are not estimated, rather band powers
$C_{\ell_{\rm eff}}$ are given that average the power spectrum through
a filter, or {\em window function}. Each theoretical model must
therefore be re-expressed in the same form before a statistical
comparison can be made. This can be done using the method of
\citet{linetal97}.

Boltzmann codes such as {\sc cmbfast} \citep{selzal96} output
theoretical power spectra in the form,
\begin{equation}
d_1({\ell})=\frac{\ell(\ell \, + \, 1)}{2\pi} \, C_{\ell}^{\rm theory}
\times {\rm normalisation} \,.
\end{equation}
Since the $C_\ell \,$s are adimensional, they are multiplied by
$T_{\rm CMB}^2 \simeq (2.725 \, {\rm K})^2$ \citep{mather99} to
express them in Kelvin,
\begin{equation}
d_2({\ell}) \, = \,  T_{\rm CMB}^2 \, d_1({\ell}) \,. 
\end{equation}

The sensitivity of each observation (denoted $i$) to a particular
$\ell$ is incorporated using the observational window function
$W_{\ell}$,
\begin{equation}
d_3(i, \, \ell) \, = \, d_2({\ell}) \times \frac{(2 \,
\ell \, +1 \, ) \, W_{\ell}^{i}}{2\, \, \ell \, (\ell \, + \, 1)} \,.
\end{equation}
The contribution from the model to the $i^{\rm th}$ observational
band-power is determined and the influence of the window function
removed,
\begin{equation}
C_{\ell_{\rm eff}}^{\rm th}(i) \, = \, \frac{\sum_{\ell =
2}^{\ell_{\rm max}} d_3(i, \, \ell)}{I(i)} \,,
\end{equation}
where $I$($i$) is the logarithmic integral of the window function, 
\begin{equation}
I(i) \, = \sum_{\ell=2}^{\ell_{\rm max}}\frac{(2 \,
\ell \, +1 \, ) \, W_{\ell}^{i}}{2\, \, \ell \, (\ell \, + \, 1)}
\,.
\end{equation}
$C_{\ell_{\rm eff}}^{\rm th}(i)$ can then be statistically compared
with the $i^{\rm th}$ band-power measurement $C_{\ell_{\rm eff}}^{\rm
obs}(i)$ given in Table~\ref{obsdat}.

The assumption that the CMB signal is a Gaussian random variable
enables analysis via a likelihood procedure. Due to the non-Gaussian
distribution of the uncertainty in the band-power measurements, an
accurate calculation of the likelihood function $L$ is
non-trivial. However, approximations to the true likelihood have been
derived \citep{bondetal00,bartlett00}.  For example, the
\citet{bondetal00} offset lognormal formalism is implemented in the
publicly available {\sc radpack} package.  Unfortunately, the
information necessary to implement this formalism has not yet been
published by all observational groups.  Therefore, in order to
statistically analyse the complete CMB observational data set, we make
the assumption that $L$ is Gaussian in $C_{\ell_{\rm eff}}$.  Then,
\begin{equation}
\label{e:chisq}
\chi^2 \equiv -2\ln{L} = \sum_i{\left(\frac{C_{\ell_{\rm eff}}^{\rm
th}(i) - C_{\ell_{\rm eff}}^{\rm obs}(i)}{\sigma^{\rm
obs}(i)}\right)^2}.
\end{equation}

The normalisation of the primordial power spectrum is not predicted by
inflationary scenarios and therefore the normalisation of the
concordance model to the full CMB observational data set is a free
parameter.  Unless we are particularly interested in the amplitude of
primordial fluctuations, we can treat the model normalisation $A$ as a
nuisance parameter.  Assuming a Gaussian likelihood, marginalisation
can be approximated numerically for the power spectrum normalisation
by computing the $\chi^2$ statistic of the concordance model for a
number of discrete steps over the normalisation range.  The
normalisation that minimises the $\chi^2$ can thereby be determined
for a particular theoretical model.

The CMB measurements have associated calibration uncertainties (see
Table~\ref{obsdat}) that allow data from the same instrument that is
calibrated using the same source to shift collectively upwards or
downwards.  The observational band-powers are multiplied by a
calibration factor $U$ that can be treated as a nuisance parameter
with a Gaussian distribution about 1.  This introduces an additional
$\chi^2$ term to Eq.~\ref{e:chisq} for each experiment that has an
associated calibration uncertainty (see Eq.~\ref{e:chisq1}).

Additionally, the BOOMERanG98, MAXIMA1 and PyV data sets have
quantified beam and/or pointing uncertainties.  The combined beam plus
pointing uncertainty for each experiment introduces an additional term
to Eq.~\ref{e:chisq} that is a function of $B$. $B$ can be treated as
a nuisance parameter, with a Gaussian distribution in $B\sigma_{b}(i)$
about 0 (see Eq.~\ref{e:chisq1}). \citet{lesgorgues01} give fitting
functions for the combined beam plus pointing uncertainty in $D^{\rm
obs}_i$ for the BOOMERanG98 and MAXIMA1 experiments; $\sigma_{b, \,
\ell} = 0.43 \times 10^{-6}\ell^2$ for BOOMERanG98 and $\sigma_{b, \,
\ell} = 10^{-6}\ell^{1.7}$ for MAXIMA1.  The 1-$\sigma$ beam
uncertainty for PyV is $\sigma_{b, \, \ell} =\exp(\pm\ell (0.425)
(0.015) (\pi/180)) - 1$ \citep{coble03}.

The nuisance parameters are incorporated into Eq.~\ref{e:chisq} to give,
\begin{equation}
\label{e:chisq1}
\chi^2 = \sum_k \left[\sum_{i=1}
^{i_{\rm max}(k)}{\left(\frac{A \,
C_{\ell_{\rm eff}}^{\rm th}(i) -
\left(U(k)+B(k)\sigma_{b}(i)\right)C_{\ell_{\rm eff}}^{\rm
obs}(i)}{\sigma(i)}\right)^2} + \left(\frac{U(k) -
1}{\sigma_{u}(k)}\right)^2 + \left(B(k)\right)^2 \right] \,,
\end{equation}
where the sum on $k$ is over the number of independent observational
data sets. 

\section{OBSERVATIONAL DATA BINNING}
\begin{figure}
\plotone{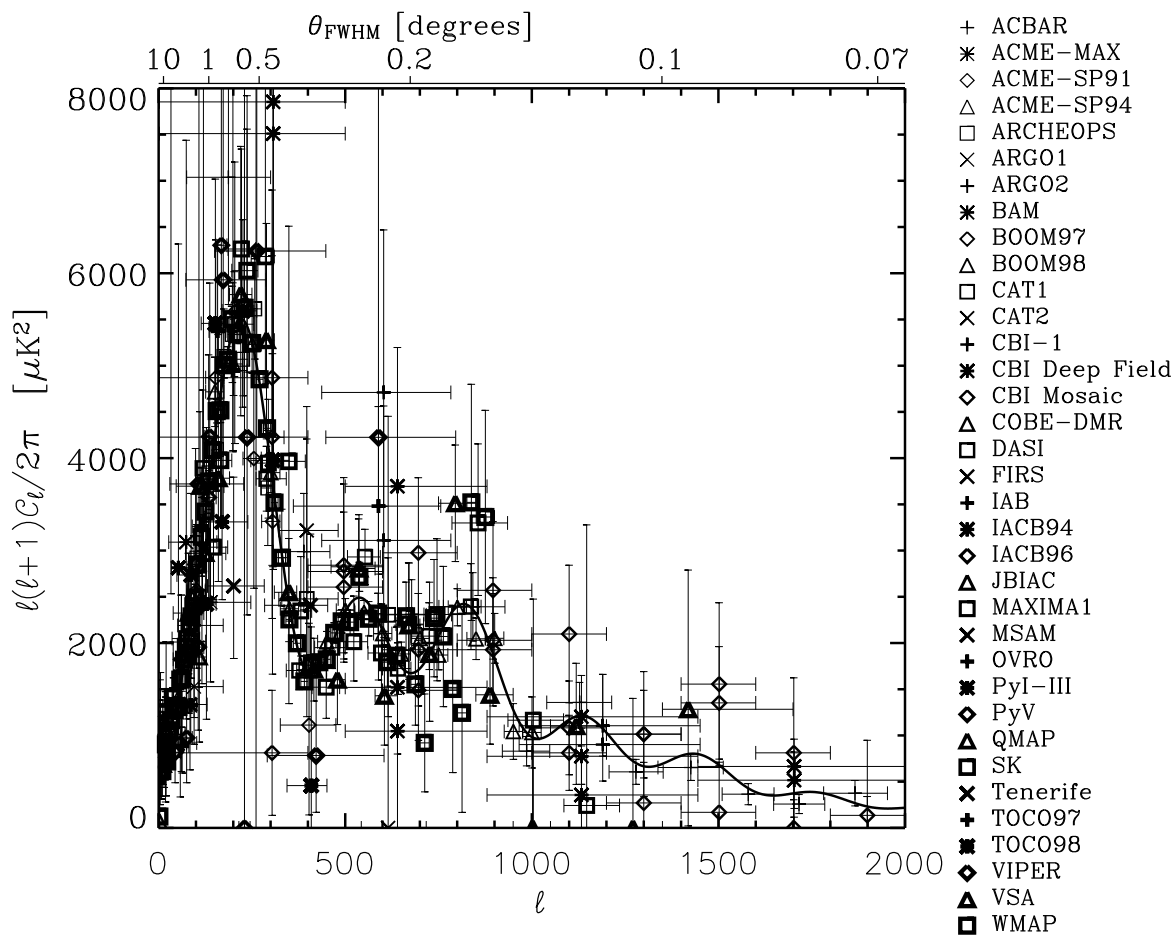}
\caption{\label{clplotcol} The concordance cosmology
(Table~\ref{concord}) normalised to the full CMB data set is plotted
with the recalibrated and, for BOOMERanG98, MAXIMA1 and PyV, beam
corrected CMB observational data (Table~\ref{obsdat}) that spans the
scales $2<\ell<2000$.}
\end{figure}

\begin{figure}
\plotone{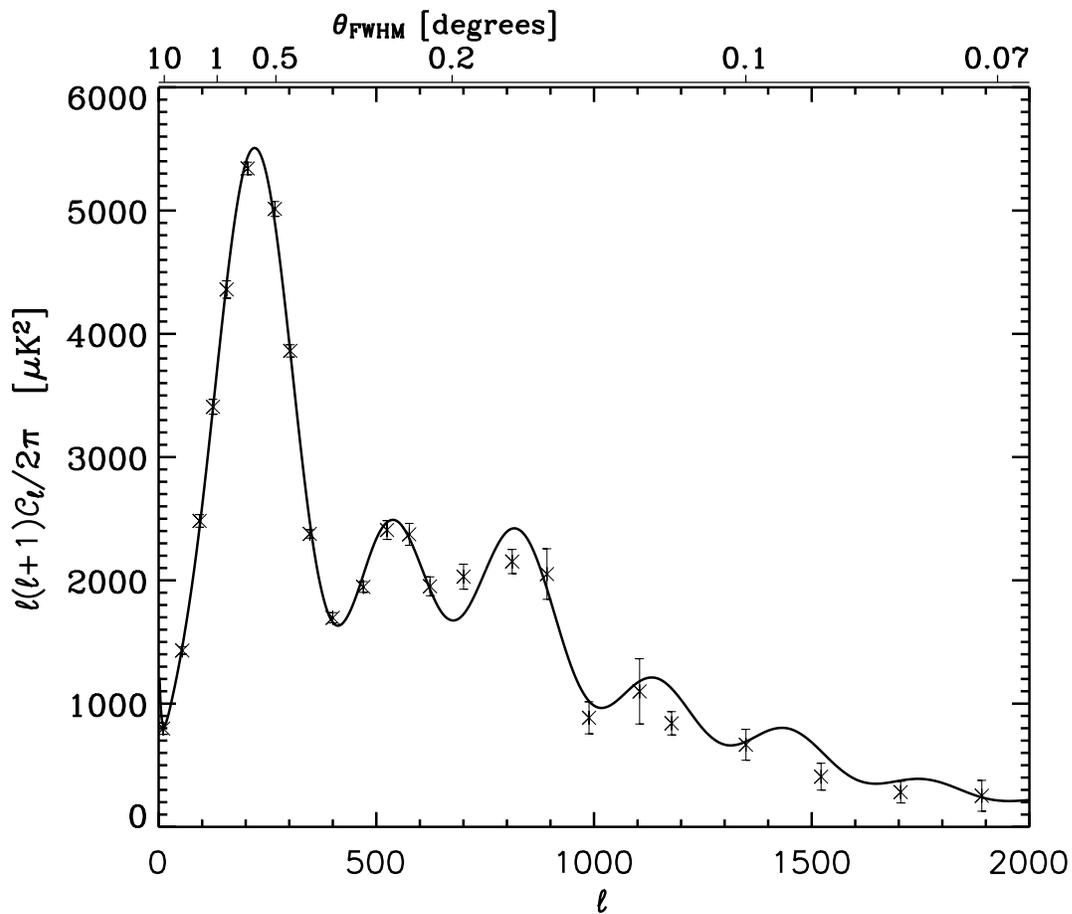}
\caption{\label{bin_clplotcol} The concordance cosmology
(Table~\ref{concord}) normalised to the full CMB data set is plotted
with the binned observational data.  The binning methodology is given
in Appendix B.  All statistical analyses detailed in this paper are
performed on the raw, unbinned data.}
\end{figure}

\begin{figure}
\plotone{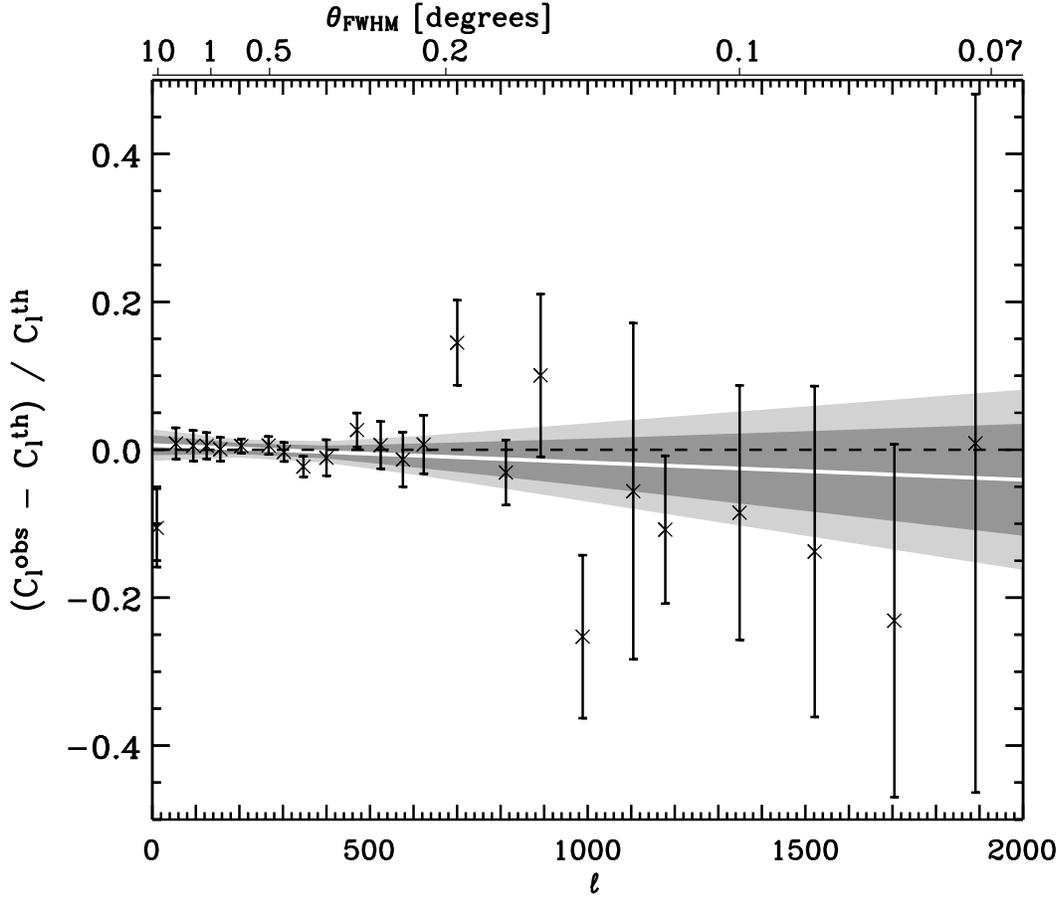}
\caption{\label{lresid}CMB data residuals plotted against $\ell$
(bottom $x$-axis) and angular scale (top $x$-axis).
Although there is little evidence for a trend in this plot,
the bins at $\ell > 900$ are predominantly low which may point to a
marginal source of systematic error or a need to slightly dampen the 
small angular scale power in the concordance model.
The $\chi^2$ per
degree of freedom for the fit of the line to the data is 
$1.05$ so the best-fit line is a slightly better fit to the data than
the zero-line that has a $\chi^2$ per degree of freedom of 
$1.06$.}
\end{figure}

\begin{figure}
\plotone{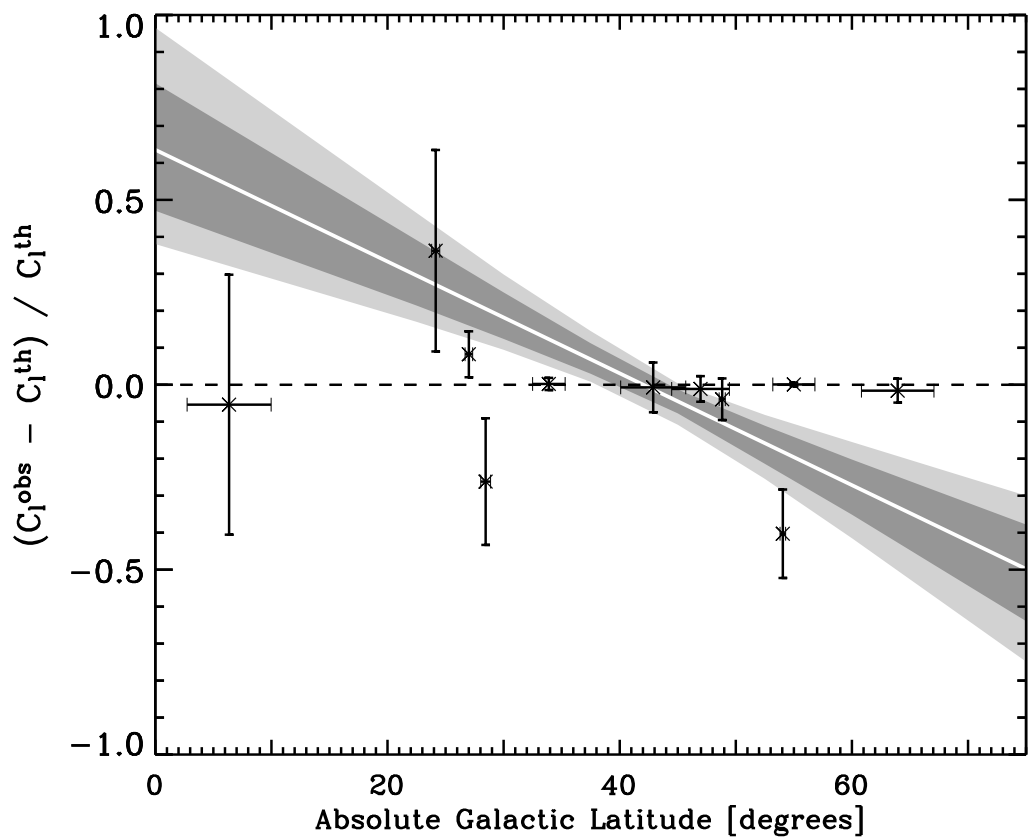} 
\caption{\label{gallatresid}CMB data residuals plotted against
absolute galactic latitude $|b|$. The fitting routine uses
uncertainties in both the $y$- and $x$-directions and assumes that the
uncertainty in $|b|$ extends to the limits of the $|b|$ range (Table
2).  The $\chi^2$ per degree of freedom for the fit of the line to the
data is $177/243 = 0.73$.  There is a more than 3-$\sigma$ trend in
this regression plot that may be indicative of a systematic error
associated with absolute galactic latitude.}
\end{figure}

\begin{figure}
\plotone{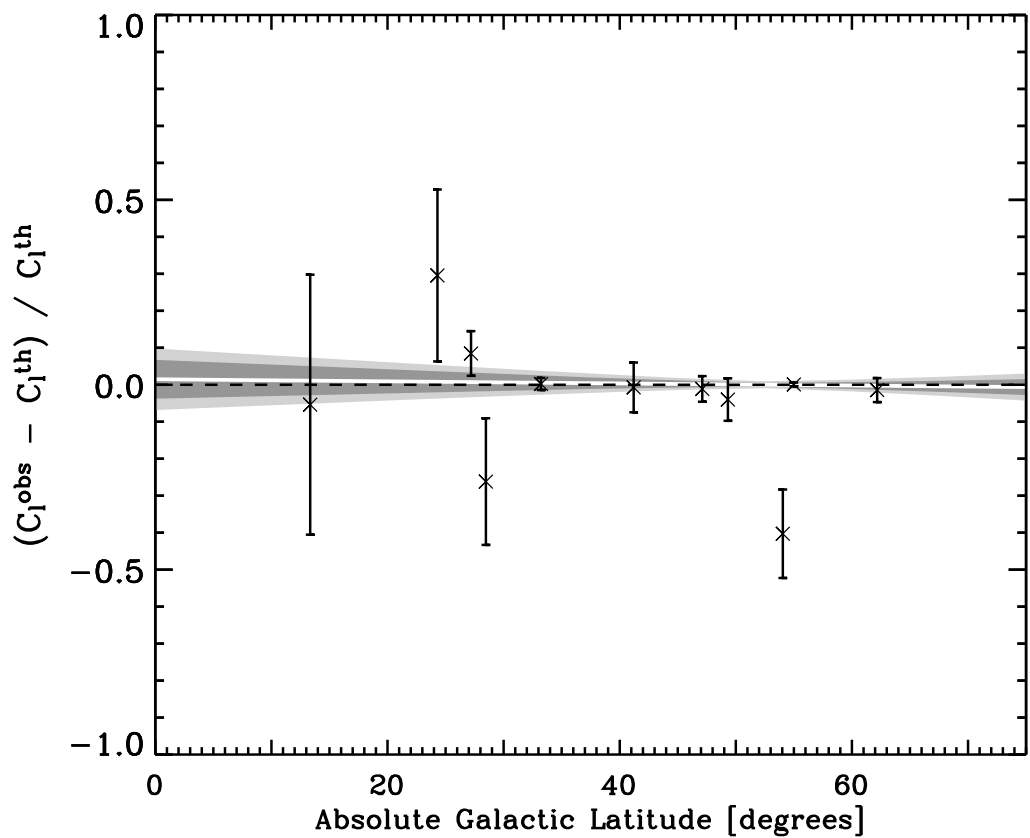}
\caption{\label{gallatresidb}CMB data residuals plotted against the
central absolute galactic latitude $|b|$, neglecting the range in
$|b|$ observed (i.e. the $x$-coordinate freedom has been removed from
the fit of Figure~\ref{gallatresid}, see \S4.2). The $\chi^2$ per
degree of freedom for the fit of the line to the data is
$258/243=1.06$.  The greater than 3-$\sigma$ trend of
Figure~\ref{gallatresid} is reduced to less than 1/2-$\sigma$ here.
The different methods used in these two plots are discussed in \S5}
\end{figure}

\begin{figure}
\plotone{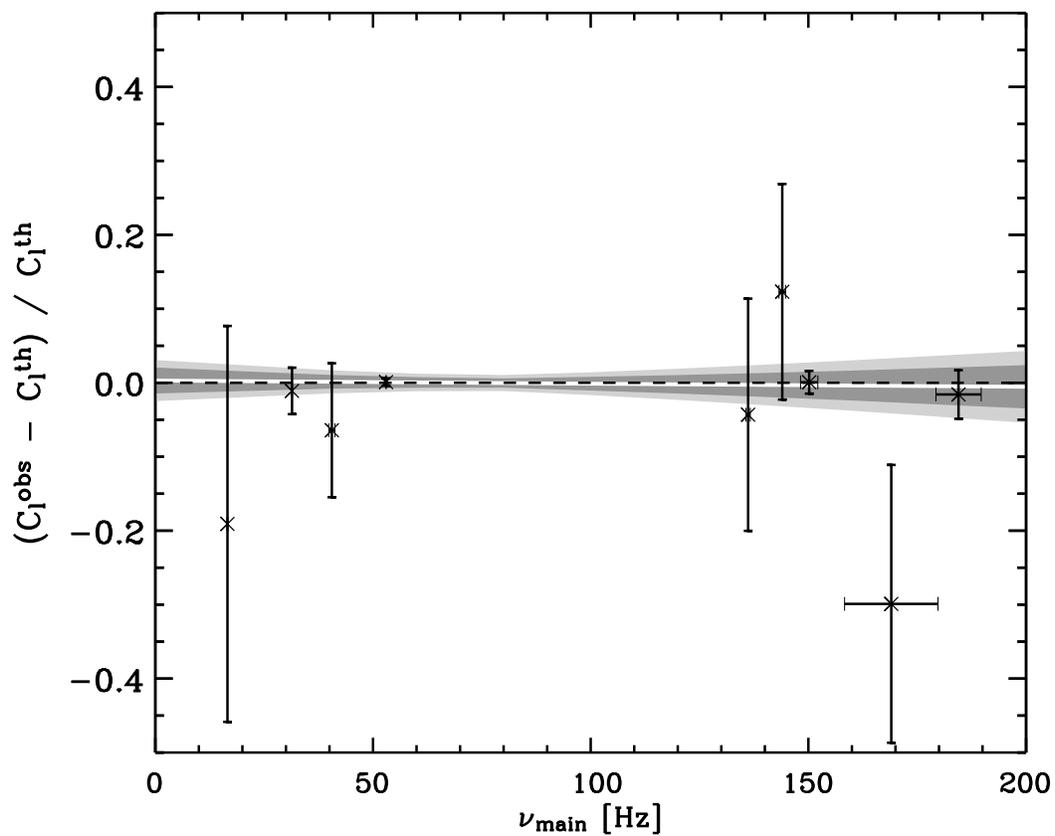} 
\caption{\label{nuresid}CMB data residuals plotted against the main
frequency of individual instruments $\nu_{\rm main}$. The fitting
routine uses uncertainties in both the $y$- and $x$-directions. The
$\chi^2$ per degree of freedom for the fit of the line to the data is
$258/243=1.06$ so the best-fit line does not improve the fit beyond
that of the zero-line. Thus, there is no evidence for a trend in this
regression plot.}
\end{figure}

\begin{figure}
\plotone{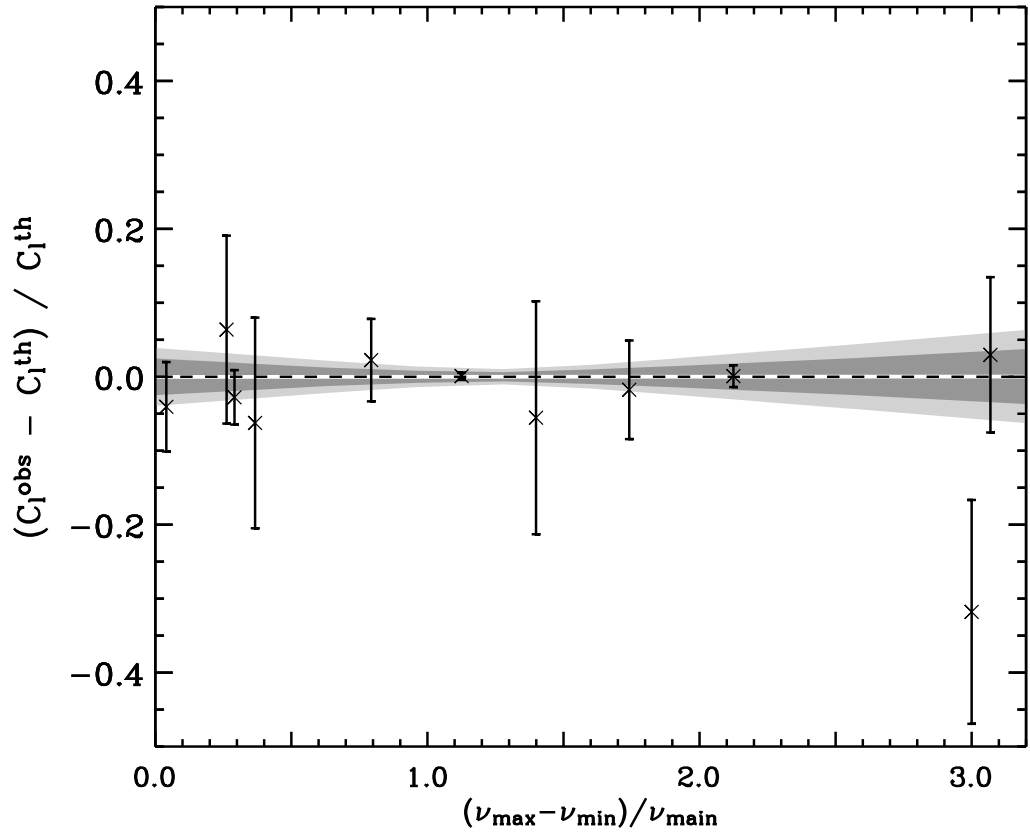}
\caption{\label{deltanuresid}CMB data residuals plotted against the
lever-arm in frequency $(\nu_{\rm max}-\nu_{\rm min})/\nu_{\rm
main}$. The $\chi^2$ per degree of freedom for the fit of the line to
the data is $258/243=1.06$ so the best-fit line does not improve the
fit beyond that of the zero-line.  Thus, there is no evidence for a
trend in this regression plot.}
\end{figure}

\begin{figure}
\plotone{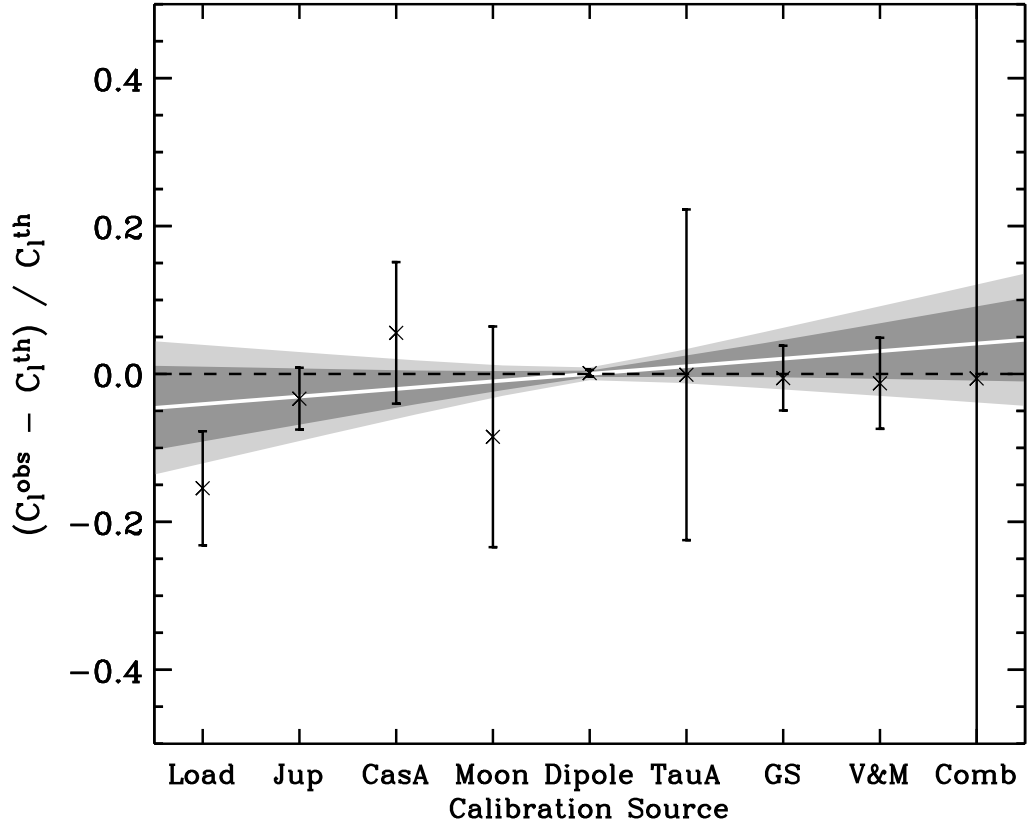} 
\caption{\label{csresid}CMB data residuals plotted against calibration
source. There are no large outliers.  The order of the calibration
sources is arbitrary so the fitting of a line serves only to verify
that the line-fitting and confidence-interval-determining code are
working as expected.  The zero-line is just inside the border of the
68\% confidence region for the best-fit line, confirming that we
should not be suspicious of trends in our residuals that are revealed
at less than the 1-$\sigma$ level.}
\end{figure}

\begin{figure}
\plotone{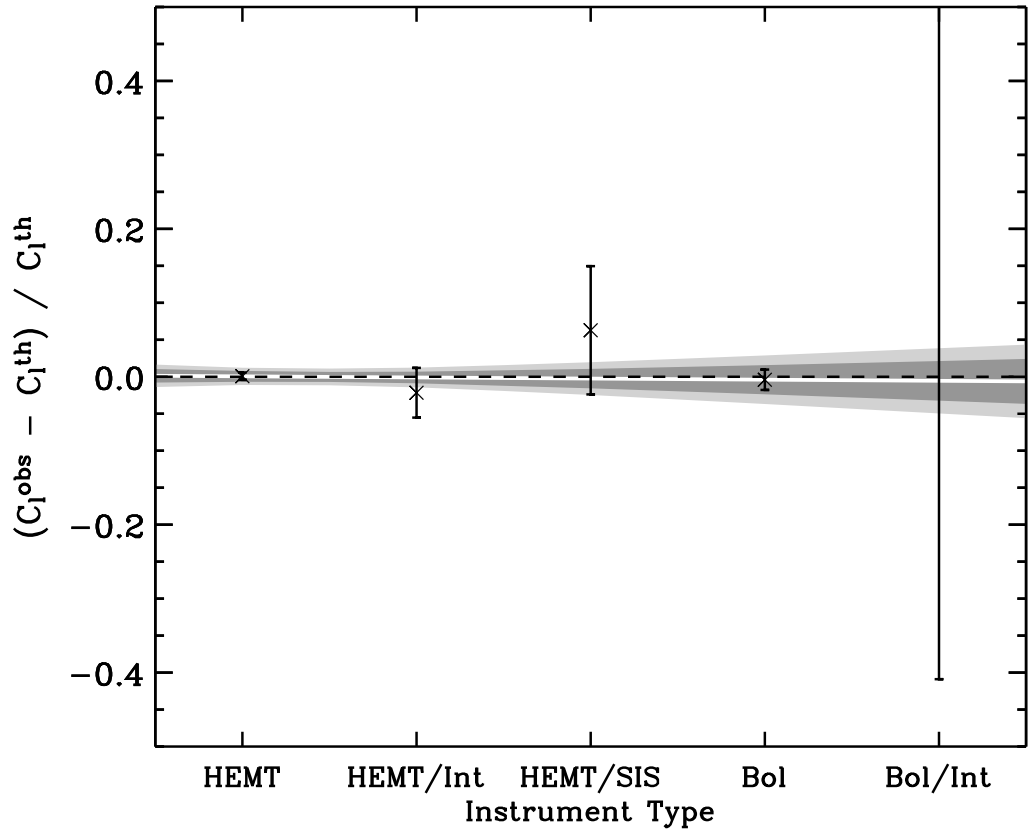} 
\caption{\label{typeresid}CMB data residuals plotted against
instrument type. There are no large outliers. As in Figure 8, 
the order of the instrument types is arbitrary so the fitting 
of a line serves only to verify that the line-fitting and 
confidence-interval-determining code are working as expected.}
\end{figure}

\begin{figure}
\plotone{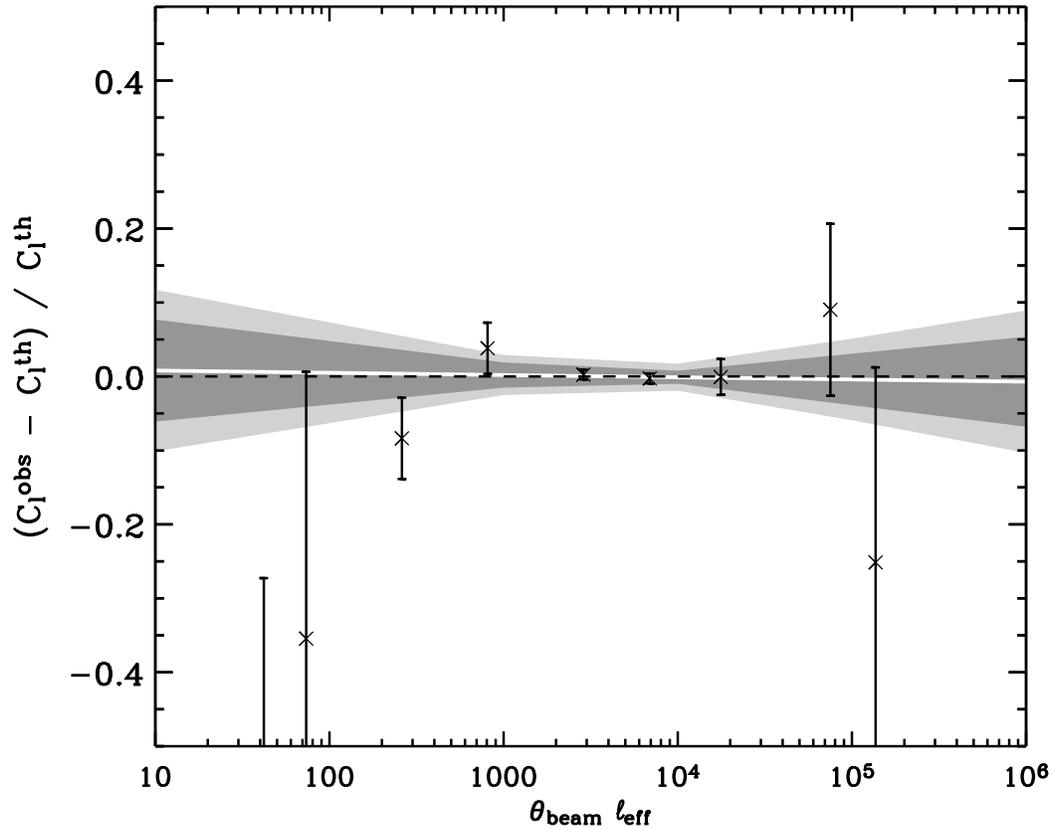}
\caption{\label{resresid}CMB data residuals plotted against
$\theta_{beam}\ell_{eff}$.  The $x$-axis is logarithmic so as to best
display the data and the residuals are examined for a linear trend
with respect to this logarithmic axis. The $\chi^2$ per degree of
freedom for the fit of the line to the data is $258/243=1.06$ so the
best-fit line does not improve the fit beyond that of the zero-line.
The motivation for this plot is to see if there are any systematics
associated with large beams sampling small scale anisotropies (right
side of plot) or with small beams sampling large scale anisotropies
(left side of plot). Small beams used to measure large angular scales
may have stability problems analogous to the problems one runs into
when trying to mosaic images together.  Although no overall linear
trend is observed, there is marginal evidence for power suppression at
$\theta_{beam}\ell_{eff} < 900$, suggesting that increased attention
should be given to band power estimates in this regime.}
\end{figure}

\begin{figure}
\plotone{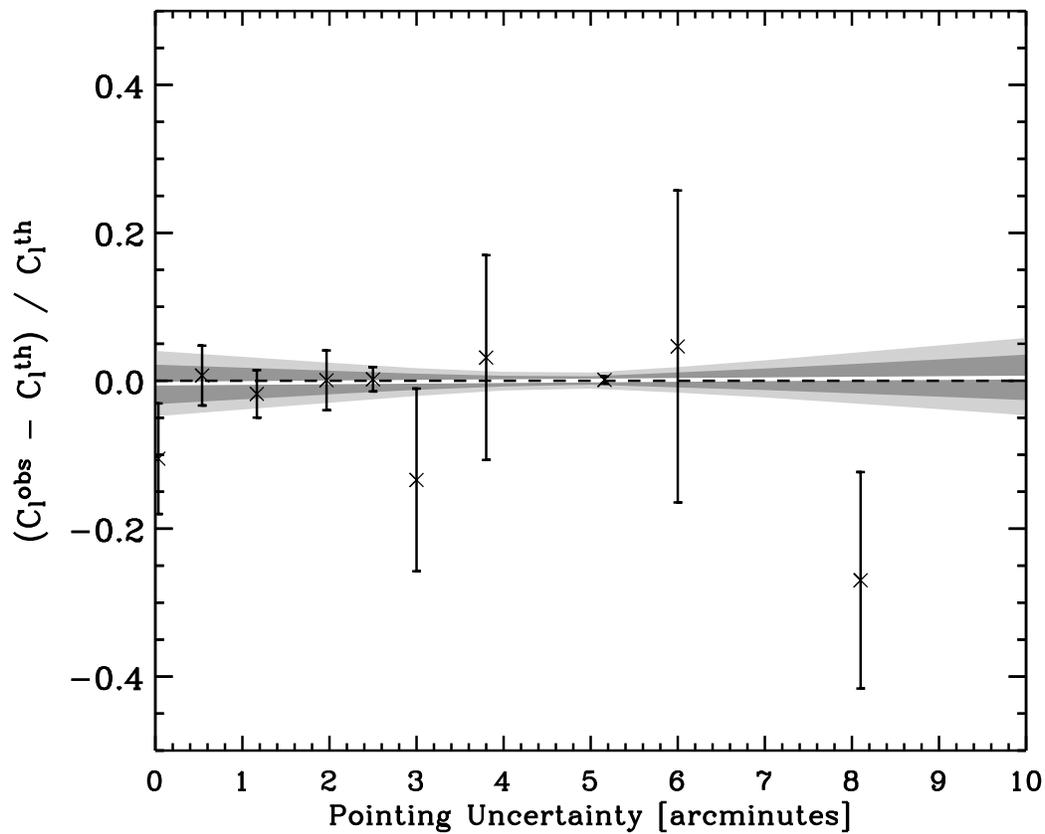}
\caption{\label{pointresid}CMB data residuals plotted against pointing
uncertainty. Six experiments do not quote pointing uncertainties so
are omitted from this analysis. With the 6 experiments omitted, there
are 239 degrees of freedom.  The $\chi^2$ per degree of freedom for
both the zero-line and the best-fitting line to the residual data is
$256/239=1.07$ so the best-fit line does not improve the fit beyond
that of the zero-line.  Thus, there is no evidence for a trend in this
regression plot.}
\end{figure}

The ever increasing number of CMB anisotropies has made data plots
such as Figure~\ref{clplotcol} difficult to interpret.  The solution
is to compress the data in some way.  Many of the more recent analyses
have chosen to concentrate on the data from just one or two
experiments, often the most recently released.  However, this not only
neglects potentially useful information, but can also unwittingly give
more weight to particular observations that may suffer from systematic
effects. We therefore choose to analyse all the available data.

One way to compress the data is to average them together into single
band-power bins in $\ell$-space.  Such an approach has been taken by a
number of authors \citep[e.g.][]{knoxpage00,jaffeetal01,wangetal02a}.
Providing that the uncertainty in the data is Gaussian and
correlations between detections are treated appropriately, narrow band
power bins can be chosen that will retain all cosmological
information.

Band-power measurements from independent observations that overlap in
the sky will be correlated to some extent.  Such correlations can only
be treated by jointly analysing the combined overlapping maps to
extract band-power estimates that are uncorrelated or have explicitly
defined correlation matrices.  This process of data compression will
wash out any systematics associated with a particular data set, so
data consistency checks are vital before this stage.  If the map data
is unavailable, the crude assumption that independent observations are
uncorrelated in space must be made.  This assumption is made in
likelihood analyses performed on the full power spectrum data set and,
since inclusion of these correlations would reduce the degrees of
freedom of an analysis, the goodness-of-fit of a particular model to
the data is better than it should be.

Some observational groups publish matrices encoding the correlations
of their individual band-power measurements.  To some extent, the
calibration uncertainties of experiments that calibrate using the same
source are also correlated. \citet{bondetal00} describe a data binning
technique that takes a lognormal noise distribution that is
approximately Gaussian and incorporates the correlation weight
matrices of individual experiments. \citet{wangetal02a} detail a
method to treat partial correlations of calibration uncertainties.
Both are useful to produce statistically meaningful data bins.

Data binning averages out any evidence for discrepancies between
independent observations and, in practice, data uncertainties are
rarely Gaussian and the information required to treat correlated data
is not always available.  So although data binning is useful for
visualisation purposes, statistical analyses of the binned
observations will generally give different results from those
performed on the raw data.

The statistical analyses detailed in this paper are performed on the
published CMB band power measurements (Table~\ref{obsdat}).  Binned
data plots are presented purely to aid the interpretation of
results. Therefore each calibrated and, for BOOMERanG98, MAXIMA1 and
PyV, beam corrected observational data point is binned assuming it to
be entirely uncorrelated.  Bin widths must be carefully chosen so that
important features of the data are not smoothed out, especially in
regions of large curvature.  For example, in the case of the power
spectrum, an unwisely chosen bin that spans an acoustic maximum will
average out the power in the bin to produce a binned data point that
misleadingly assigns less power to the peak.

The contribution from the $i$th observational measurement
($x_i\pm\sigma_{x, \, i}$, $y_i\pm\sigma_{y,\, i}$) to a binned point
($x_b\pm\sigma_{x, \, b}$, $y_b\pm\sigma_{y,\, b}$) is inverse
variance weighted,
\begin{eqnarray}
x_b \, \pm \, \sigma_{x, \, b} \, = \, \frac{\sum_i x_i \, \sigma_{x, \, i}^{-2}}{\sum_i\sigma_{x, \, i}^{-2}} \, \pm \, \sqrt{\frac{1}{\sum_i\sigma_{x, \, i}^{-2}}} \,, \\
y_b \, \pm \, \sigma_{y, \, b}\,  = \, \frac{\sum_i y_i \, \sigma_{y, \, i}^{-2}}{\sum_i\sigma_{y, \, i}^{-2}} \, \pm \, \sqrt{\frac{1}{\sum_i\sigma_{y, \, i}^{-2}}} \,.
\end{eqnarray}
If quoted error bars are asymmetric, a first guess for the binned data
point is obtained by averaging the uncertainties.  A more accurate
estimate can then be converged upon by iterating over the binning
routine, inserting the positive variance for measurements that are
below the bin averaged point and negative variance for those that are
above.

All the data from a particular experiment will be measured using the
same instrument and therefore can be binned together for the purpose
of visualising any trends in the data residuals with respect to the
instrument design. When data from the same experiment is placed in the
same bin, the variance of the resultant binned data point can be
easily adjusted to account for any correlated calibration uncertainty
associated with the observational data. A degree of freedom is reduced
for each independent calibration uncertainty. This effectively
tightens the constraints on the binned data.

For example, if $n$ calibrated data points in a bin have equal
variance $\sigma_y$ and an entirely correlated calibration
uncertainty, they share $n-1$ degrees of freedom and their
contribution to the variance of the binned data point is then
$\sigma_y/\sqrt{n-1}$. When each of the $n$ data points have different
variances, their contribution to the uncertainty in the binned data
point is given by,
\begin{equation}
\sigma_{y, \, b} \, = \, \frac{1}{\sqrt{\sum_j\sigma_{y, \, j}^{-2} - \left(n \, / \, \sum_j\sigma_{y, \, j}\right)^2}} \,.
\end{equation}
This method of binning is employed when appropriate to produce the
plotted residual data bins.

\clearpage

\clearpage

\clearpage

\clearpage

\clearpage

\clearpage

\clearpage

\clearpage

\clearpage

\clearpage

\clearpage

\clearpage
\begin{deluxetable}{lccccccccccc} 
\rotate
\tabletypesize{\tiny}
\tablecolumns{12} 
\tablewidth{0pc}
\setlength{\tabcolsep}{0.04in}
\tablecaption{\label{concord} 1-$\sigma$ cosmological parameter constraints from five analyses.}
\tablehead{ 
\colhead{}    &  \multicolumn{2}{c}{\cite{efstetal02}} & \multicolumn{2}{c}{\cite{sievers02}} & \multicolumn{2}{c}{\cite{lewisbridle02}} & \multicolumn{2}{c}{\cite{wangetal02b}} & \multicolumn{2}{c}{\cite{spergel03}}  & $\Lambda$CDM concordance \\
\colhead{} & \colhead{CMB alone}   & \colhead{+2dFGRS+BBN}   & \colhead{CMB alone}   & \colhead{+priors$^a$} & \colhead{CMB+priors$^b$} &\colhead{+2dFGRS} & \colhead{CMB+flat prior}   & \colhead{+2dFGRS} & \colhead{CMB(TT+TE)+flat prior} & \colhead{+2dFGRS+Ly$\alpha$} &\colhead{} }
\startdata
$\Omega_k$       &$-0.04^{+.05}_{-.32}$  &$-0.013^{+.027}_{-.019}$&$-0.05\pm .05$         & (0.00)                & (0.00)           & (0.00)           & (0.00) 		& (0.00)           &(0.00)   &(0.00)  & 0.0    \\
$\Omega_\Lambda$ &$0.43^{+.23}$          &$0.73\pm .04$            &$0.54^{+.12}_{-.13}$   &$0.70^{+.02}_{-.03}$   & $0.72\pm 0.06$   & $0.71 \pm 0.04$  &$0.71 \pm 0.11$   & $0.72 \pm 0.09$   &$0.76^{+.05}_{-.06}$    &$0.74^{+.03}_{-.04}$   &0.74 \\
$\omega_b$       &$0.020^{+.013}_{-.002}$&$0.020\pm .001$          &$0.023\pm .003$        &$0.024^{+.002}_{-.003}$& $0.022 \pm .001$ & $0.022 \pm .001$ & 0.023$\pm$.003 	& 0.024$\pm$.003   &$0.023\pm .001$ &$0.0226 \pm .0008$ &0.0226\\
$\omega_c$       &$0.13^{+.03}_{-.05}$   &$0.10^{+.02}_{-.01}$     & $0.13^{+.03}_{-.02}$  &$0.12^{+.01}_{-.01}$   & -                & -                & 0.112$\pm$.014 	& 0.115$\pm$.013   &$0.11^{+.06}_{-.04}$    &$0.11\pm .03$  &0.11\\
$\omega_d$       &-                      &-                        &-                      &-                      & 0.099$\pm$.014   & 0.106$\pm$.010   & - 		& -                & - & - &- \\
$f_{\nu}$	    &-                      &-                        &-                      &-                      &$<0.10$           &$<0.04$           & - 		& -                & - & - &0 \\
$n_s$            &$0.96^{+.27}_{-.04}$ &$1.04^{+.06}_{-.05}$       &$1.02^{+.06}_{-.07}$   &$1.04^{+.05}_{-.06}$   & 1.02$\pm$.05     &1.03$\pm$.05      &0.99$\pm$.06 	& 0.99$\pm$.04     &$0.97 \pm .03$    &$0.96 \pm .02 $   &0.96 \\
$\tau$           &$<0.25$                &$<0.25$                  &$0.16^{+.18}_{-.13}$   &$0.13^{+.13}_{-.10}$   & -                & -                & $0.04^{+0.06}$    &$0.06 \pm .03$    &$0.14^{+.07}_{-.06}$                 &$0.12^{+.06}_{-.05}$                &0.12  \\
$\Omega_m h$     &-                      &$0.19 \pm .02$           &-                      &-                      & 0.18$\pm$.03     &0.19$\pm$.02      &- 		&-                         &-                       &-                      &0.185\\
$h$              &-                      &$0.66^{+.09}_{-.03}$     &$0.55^{+.09}_{-.09}$   &$0.69^{+.02}_{-.02}$   & 0.67$\pm$.05     & 0.66$\pm$.03     & 0.71$\pm$.13 	& 0.73$\pm$.11     &$0.73 \pm .05$    &$0.72 \pm .03$   & 0.72\\
\enddata 
\tablenotetext{a}{The priors used in this analysis are a flat prior $\Omega_k=0$ in accordance with the predictions of the simplest inflationary scenarios, a large scale structure prior that involves a constraint on the amplitude $\sigma_8^2$ and the shape of the matter power spectrum, the HKP prior for $h$ and the SNIa priors.}
\tablenotetext{b}{The priors used in this analysis are a flat prior $\Omega_k=0$ in accordance with the predictions of the simplest inflationary scenarios, the BBN prior for $\omega_b$, the HKP prior for $h$ and the SNIa priors.}
\end{deluxetable}

\clearpage
\begin{deluxetable}{lccccccc}
\tabletypesize{\tiny}
\tablecolumns{8} 
\tablewidth{0pc} 
\tablecaption{\label{obsdat} The current compilation of CMB observational data from $\ell=2$ to $\ell=2000$.} 
\tablehead{ 
\colhead{Experiment} & \colhead{Ref.}   & \colhead{$\ell_{{\rm eff}}$}    & \colhead{$\ell_{{\rm min}}$} & 
\colhead{$\ell_{{\rm max}}$}    & \colhead{$C_{\ell_{\rm eff}}^{\rm obs}\pm \sigma^{\rm obs}$}   & \colhead{$\sigma_{u}^{a}$}    & \colhead{ Publication Date}\\
\colhead{} & \colhead{}   & \colhead{}    & \colhead{} & 
\colhead{}    & \colhead{ $(\mu$K)}   & \colhead{(\%)}    & \colhead{(yrs)}}
\startdata 
       ACBAR &  [1] &  187.0 &   75.0 &  300.0 & $  6767.0_{-  1323.0}^{+  1323.0}$ & 20.0 & 2002.9 \\
       ACBAR &  [1] &  389.0 &  307.0 &  459.0 & $  2874.0_{-   605.0}^{+   605.0}$ & 20.0 & 2002.9 \\
       ACBAR &  [1] &  536.0 &  462.0 &  602.0 & $  2716.0_{-   498.0}^{+   498.0}$ & 20.0 & 2002.9 \\
       ACBAR &  [1] &  678.0 &  615.0 &  744.0 & $  2222.0_{-   360.0}^{+   360.0}$ & 20.0 & 2002.9 \\
       ACBAR &  [1] &  842.0 &  751.0 &  928.0 & $  2300.0_{-   355.0}^{+   355.0}$ & 20.0 & 2002.9 \\
       ACBAR &  [1] &  986.0 &  921.0 & 1048.0 & $   798.0_{-   153.0}^{+   153.0}$ & 20.0 & 2002.9 \\
       ACBAR &  [1] & 1128.0 & 1040.0 & 1214.0 & $  1305.0_{-   208.0}^{+   208.0}$ & 20.0 & 2002.9 \\
       ACBAR &  [1] & 1279.0 & 1207.0 & 1352.0 & $   583.0_{-   130.0}^{+   130.0}$ & 20.0 & 2002.9 \\
       ACBAR &  [1] & 1426.0 & 1338.0 & 1513.0 & $   628.0_{-   134.0}^{+   134.0}$ & 20.0 & 2002.9 \\
       ACBAR &  [1] & 1580.0 & 1510.0 & 1649.0 & $   351.0_{-   110.0}^{+   110.0}$ & 20.0 & 2002.9 \\
       ACBAR &  [1] & 1716.0 & 1648.0 & 1785.0 & $   248.0_{-    99.0}^{+    99.0}$ & 20.0 & 2002.9 \\
       ACBAR &  [1] & 1866.0 & 1782.0 & 1953.0 & $   361.0_{-   132.0}^{+   132.0}$ & 20.0 & 2002.9 \\
    ACME-MAX &  [2] &  139.0 &   72.0 &  247.0 & $  2440.4_{-   861.4}^{+   964.1}$ & -  &1996.7 \\
   ACME-SP91 &  [3] &   61.0 &   30.0 &  102.0 & $   918.1_{-   308.0}^{+   618.5}$ & -  &1995.3 \\
   ACME-SP94 &  [3] &   61.0 &   30.0 &  102.0 & $  1317.7_{-   473.6}^{+  1201.6}$ & -  &1995.3 \\
    ARCHEOPS &  [4] &   18.5 &   15.0 &   22.0 & $   789.0_{-   537.0}^{+   537.0}$ & 14.0 & 2002.8 \\
    ARCHEOPS &  [4] &   28.5 &   22.0 &   35.0 & $   936.0_{-   230.0}^{+   230.0}$ & 14.0 & 2002.8 \\
    ARCHEOPS &  [4] &   40.0 &   35.0 &   45.0 & $  1198.0_{-   262.0}^{+   262.0}$ & 14.0 & 2002.8 \\
    ARCHEOPS &  [4] &   52.5 &   45.0 &   60.0 & $   912.0_{-   224.0}^{+   224.0}$ & 14.0 & 2002.8 \\
    ARCHEOPS &  [4] &   70.0 &   60.0 &   80.0 & $  1596.0_{-   224.0}^{+   224.0}$ & 14.0 & 2002.8 \\
    ARCHEOPS &  [4] &   87.5 &   80.0 &   95.0 & $  1954.0_{-   280.0}^{+   280.0}$ & 14.0 & 2002.8 \\
    ARCHEOPS &  [4] &  102.5 &   95.0 &  110.0 & $  2625.0_{-   325.0}^{+   325.0}$ & 14.0 & 2002.8 \\
    ARCHEOPS &  [4] &  117.5 &  110.0 &  125.0 & $  2681.0_{-   364.0}^{+   364.0}$ & 14.0 & 2002.8 \\
    ARCHEOPS &  [4] &  135.0 &  125.0 &  145.0 & $  3454.0_{-   358.0}^{+   358.0}$ & 14.0 & 2002.8 \\
    ARCHEOPS &  [4] &  155.0 &  145.0 &  165.0 & $  3681.0_{-   396.0}^{+   396.0}$ & 14.0 & 2002.8 \\
    ARCHEOPS &  [4] &  175.0 &  165.0 &  185.0 & $  4586.0_{-   462.0}^{+   462.0}$ & 14.0 & 2002.8 \\
    ARCHEOPS &  [4] &  197.5 &  185.0 &  210.0 & $  4801.0_{-   469.0}^{+   469.0}$ & 14.0 & 2002.8 \\
    ARCHEOPS &  [4] &  225.0 &  210.0 &  240.0 & $  4559.0_{-   467.0}^{+   467.0}$ & 14.0 & 2002.8 \\
    ARCHEOPS &  [4] &  257.5 &  240.0 &  275.0 & $  5049.0_{-   488.0}^{+   488.0}$ & 14.0 & 2002.8 \\
    ARCHEOPS &  [4] &  292.5 &  275.0 &  310.0 & $  3307.0_{-   560.0}^{+   560.0}$ & 14.0 & 2002.8 \\
    ARCHEOPS &  [4] &  330.0 &  310.0 &  350.0 & $  2629.0_{-   471.0}^{+   471.0}$ & 14.0 & 2002.8 \\
       ARGO1 &  [5] &   95.0 &   51.0 &  173.0 & $  1528.8_{-   625.4}^{+   742.5}$ & -  &1994.1 \\
       ARGO2 &  [6] &   95.0 &   51.0 &  173.0 & $  2190.2_{-   996.6}^{+   970.8}$ & -  &1996.4 \\
         BAM &  [7] &   74.0 &   27.0 &  156.0 & $  3091.4_{-  1912.7}^{+  4347.3}$ & -  &1997.2 \\
      BOOM97 &  [8] &   58.0 &   25.0 &   75.0 & $   850.0_{-   540.0}^{+   900.0}$ & 16.0 & 2000.4 \\
      BOOM97 &  [8] &  102.0 &   76.0 &  125.0 & $  2380.0_{-   780.0}^{+   990.0}$ & 16.0 & 2000.4 \\
      BOOM97 &  [8] &  153.0 &  126.0 &  175.0 & $  4510.0_{-  1140.0}^{+  1380.0}$ & 16.0 & 2000.4 \\
      BOOM97 &  [8] &  204.0 &  204.0 &  255.0 & $  5170.0_{-  1320.0}^{+  1500.0}$ & 16.0 & 2000.4 \\
      BOOM97 &  [8] &  255.0 &  226.0 &  275.0 & $  3700.0_{-  1300.0}^{+  1500.0}$ & 16.0 & 2000.4 \\
      BOOM97 &  [8] &  305.0 &  276.0 &  325.0 & $  3070.0_{-  1530.0}^{+  1680.0}$ & 16.0 & 2000.4 \\
      BOOM97 &  [8] &  403.0 &  326.0 &  475.0 & $  1030.0_{-   900.0}^{+  1020.0}$ & 16.0 & 2000.4 \\
      BOOM98 &  [9] &   50.5 &   26.0 &   75.0 & $  1423.0_{-   313.0}^{+   313.0}$ & 20.0 & 2002.9 \\
      BOOM98 &  [9] &  100.5 &   76.0 &  125.0 & $  2609.0_{-   279.0}^{+   279.0}$ & 20.0 & 2002.9 \\
      BOOM98 &  [9] &  150.5 &  126.0 &  175.0 & $  4823.0_{-   384.0}^{+   384.0}$ & 20.0 & 2002.9 \\
      BOOM98 &  [9] &  200.5 &  176.0 &  225.0 & $  5139.0_{-   349.0}^{+   349.0}$ & 20.0 & 2002.9 \\
      BOOM98 &  [9] &  250.5 &  226.0 &  275.0 & $  5365.0_{-   321.0}^{+   321.0}$ & 20.0 & 2002.9 \\
      BOOM98 &  [9] &  300.5 &  276.0 &  325.0 & $  3953.0_{-   222.0}^{+   222.0}$ & 20.0 & 2002.9 \\
      BOOM98 &  [9] &  350.5 &  326.0 &  375.0 & $  2445.0_{-   137.0}^{+   137.0}$ & 20.0 & 2002.9 \\
      BOOM98 &  [9] &  400.5 &  376.0 &  425.0 & $  1822.0_{-   105.0}^{+   105.0}$ & 20.0 & 2002.9 \\
      BOOM98 &  [9] &  450.5 &  426.0 &  475.0 & $  2092.0_{-   116.0}^{+   116.0}$ & 20.0 & 2002.9 \\
      BOOM98 &  [9] &  500.5 &  476.0 &  525.0 & $  2456.0_{-   132.0}^{+   132.0}$ & 20.0 & 2002.9 \\
      BOOM98 &  [9] &  550.5 &  526.0 &  575.0 & $  2444.0_{-   135.0}^{+   135.0}$ & 20.0 & 2002.9 \\
      BOOM98 &  [9] &  600.5 &  576.0 &  625.0 & $  2216.0_{-   133.0}^{+   133.0}$ & 20.0 & 2002.9 \\
      BOOM98 &  [9] &  650.5 &  626.0 &  675.0 & $  1994.0_{-   136.0}^{+   136.0}$ & 20.0 & 2002.9 \\
      BOOM98 &  [9] &  700.5 &  676.0 &  725.0 & $  2186.0_{-   157.0}^{+   157.0}$ & 20.0 & 2002.9 \\
      BOOM98 &  [9] &  750.5 &  726.0 &  775.0 & $  2008.0_{-   172.0}^{+   172.0}$ & 20.0 & 2002.9 \\
      BOOM98 &  [9] &  800.5 &  776.0 &  825.0 & $  2581.0_{-   217.0}^{+   217.0}$ & 20.0 & 2002.9 \\
      BOOM98 &  [9] &  850.5 &  826.0 &  875.0 & $  2229.0_{-   245.0}^{+   245.0}$ & 20.0 & 2002.9 \\
      BOOM98 &  [9] &  900.5 &  876.0 &  925.0 & $  2253.0_{-   296.0}^{+   296.0}$ & 20.0 & 2002.9 \\
      BOOM98 &  [9] &  950.5 &  926.0 &  975.0 & $  1156.0_{-   334.0}^{+   334.0}$ & 20.0 & 2002.9 \\
      BOOM98 &  [9] & 1000.5 &  976.0 & 1025.0 & $  1155.0_{-   430.0}^{+   430.0}$ & 20.0 & 2002.9 \\
        CAT1 & [10] &  397.0 &  322.0 &  481.0 & $  2580.6_{-  1327.5}^{+  1801.8}$ & 20.0 & 1996.3 \\
        CAT1 & [10] &  615.0 &  543.0 &  717.0 & $  2401.0_{-  1147.8}^{+  2236.6}$ & 20.0 & 1996.3 \\
        CAT2 & [11] &  397.0 &  322.0 &  481.0 & $  3283.3_{-  1373.6}^{+  1367.9}$ & 20.0 & 1999.8 \\
        CAT2 & [11] &  615.0 &  543.0 &  717.0 & $     0.0_{-     0.0}^{+  2981.0}$ & 20.0 & 1999.8 \\
      CBI-1a & [12] &  603.0 &  437.0 &  783.0 & $  4096.0_{-  1071.0}^{+  1529.0}$ & 10.0 & 2001.2 \\
      CBI-1a & [12] & 1190.0 &  966.0 & 1451.0 & $   961.0_{-   285.0}^{+   483.0}$ & 10.0 & 2001.2 \\
      CBI-1b & [12] &  603.0 &  437.0 &  783.0 & $  2704.0_{-   855.0}^{+  1265.0}$ & 10.0 & 2001.2 \\
      CBI-1b & [12] & 1190.0 &  966.0 & 1451.0 & $   784.0_{-   343.0}^{+   660.0}$ & 10.0 & 2001.2 \\
     CBI-D8h & [13] &  307.0 &    2.0 &  500.0 & $  6531.0_{-  4100.0}^{+  4100.0}$ & 10.0 & 2002.4 \\
     CBI-D8h & [13] &  640.0 &  500.0 &  880.0 & $  3213.0_{-  1306.0}^{+  1306.0}$ & 10.0 & 2002.4 \\
     CBI-D8h & [13] & 1133.0 &  880.0 & 1445.0 & $  1045.0_{-   389.0}^{+   389.0}$ & 10.0 & 2002.4 \\
     CBI-D8h & [13] & 1703.0 & 1445.0 & 2010.0 & $   449.0_{-   266.0}^{+   266.0}$ & 10.0 & 2002.4 \\
    CBI-D14h & [13] &  307.0 &    2.0 &  500.0 & $  8381.0_{-  5274.0}^{+  5274.0}$ & 10.0 & 2002.4 \\
    CBI-D14h & [13] &  640.0 &  500.0 &  880.0 & $   910.0_{-   983.0}^{+   983.0}$ & 10.0 & 2002.4 \\
    CBI-D14h & [13] & 1133.0 &  880.0 & 1445.0 & $   310.0_{-   461.0}^{+   461.0}$ & 10.0 & 2002.4 \\
    CBI-D14h & [13] & 1703.0 & 1445.0 & 2010.0 & $  -415.0_{-   516.0}^{+   516.0}$ & 10.0 & 2002.4 \\
    CBI-D20h & [13] &  307.0 &    2.0 &  500.0 & $  6829.0_{-  3540.0}^{+  3540.0}$ & 10.0 & 2002.4 \\
    CBI-D20h & [13] &  640.0 &  500.0 &  880.0 & $  1322.0_{-   627.0}^{+   627.0}$ & 10.0 & 2002.4 \\
    CBI-D20h & [13] & 1133.0 &  880.0 & 1445.0 & $   674.0_{-   264.0}^{+   264.0}$ & 10.0 & 2002.4 \\
    CBI-D20h & [13] & 1703.0 & 1445.0 & 2010.0 & $   578.0_{-   257.0}^{+   257.0}$ & 10.0 & 2002.4 \\
     CBI-M2h & [14] &  304.0 &    2.0 &  400.0 & $   706.0_{-   588.0}^{+   588.0}$ & 10.0 & 2002.4 \\
     CBI-M2h & [14] &  496.0 &  400.0 &  600.0 & $  2265.0_{-   706.0}^{+   706.0}$ & 10.0 & 2002.4 \\
     CBI-M2h & [14] &  696.0 &  600.0 &  800.0 & $  1676.0_{-   529.0}^{+   529.0}$ & 10.0 & 2002.4 \\
     CBI-M2h & [14] &  896.0 &  800.0 & 1000.0 & $  2235.0_{-   647.0}^{+   647.0}$ & 10.0 & 2002.4 \\
     CBI-M2h & [14] & 1100.0 & 1000.0 & 1200.0 & $   941.0_{-   441.0}^{+   441.0}$ & 10.0 & 2002.4 \\
     CBI-M2h & [14] & 1300.0 & 1200.0 & 1400.0 & $   235.0_{-   382.0}^{+   382.0}$ & 10.0 & 2002.4 \\
     CBI-M2h & [14] & 1502.0 & 1400.0 & 1600.0 & $   147.0_{-   471.0}^{+   471.0}$ & 10.0 & 2002.4 \\
     CBI-M2h & [14] & 1702.0 & 1600.0 & 1800.0 & $     0.0_{-   471.0}^{+   471.0}$ & 10.0 & 2002.4 \\
     CBI-M2h & [14] & 1899.0 & 1800.0 & 2000.0 & $  -176.0_{-   471.0}^{+   471.0}$ & 10.0 & 2002.4 \\
    CBI-M14h & [14] &  304.0 &    2.0 &  400.0 & $  4235.0_{-  1765.0}^{+  1765.0}$ & 10.0 & 2002.4 \\
    CBI-M14h & [14] &  496.0 &  400.0 &  600.0 & $  2471.0_{-   824.0}^{+   824.0}$ & 10.0 & 2002.4 \\
    CBI-M14h & [14] &  696.0 &  600.0 &  800.0 & $  2588.0_{-   706.0}^{+   706.0}$ & 10.0 & 2002.4 \\
    CBI-M14h & [14] &  896.0 &  800.0 & 1000.0 & $  1676.0_{-   529.0}^{+   529.0}$ & 10.0 & 2002.4 \\
    CBI-M14h & [14] & 1100.0 & 1000.0 & 1200.0 & $   706.0_{-   353.0}^{+   353.0}$ & 10.0 & 2002.4 \\
    CBI-M14h & [14] & 1300.0 & 1200.0 & 1400.0 & $   882.0_{-   412.0}^{+   412.0}$ & 10.0 & 2002.4 \\
    CBI-M14h & [14] & 1502.0 & 1400.0 & 1600.0 & $  1176.0_{-   529.0}^{+   529.0}$ & 10.0 & 2002.4 \\
    CBI-M14h & [14] & 1702.0 & 1600.0 & 1800.0 & $   -88.0_{-   353.0}^{+   353.0}$ & 10.0 & 2002.4 \\
    CBI-M14h & [14] & 1899.0 & 1800.0 & 2000.0 & $  -471.0_{-   353.0}^{+   353.0}$ & 10.0 & 2002.4 \\
    CBI-M20h & [14] &  304.0 &    2.0 &  400.0 & $  3676.0_{-  1706.0}^{+  1706.0}$ & 10.0 & 2002.4 \\
    CBI-M20h & [14] &  496.0 &  400.0 &  600.0 & $  2412.0_{-   823.0}^{+   823.0}$ & 10.0 & 2002.4 \\
    CBI-M20h & [14] &  696.0 &  600.0 &  800.0 & $  1294.0_{-   471.0}^{+   471.0}$ & 10.0 & 2002.4 \\
    CBI-M20h & [14] &  896.0 &  800.0 & 1000.0 & $  1765.0_{-   588.0}^{+   588.0}$ & 10.0 & 2002.4 \\
    CBI-M20h & [14] & 1100.0 & 1000.0 & 1200.0 & $  1824.0_{-   647.0}^{+   647.0}$ & 10.0 & 2002.4 \\
    CBI-M20h & [14] & 1300.0 & 1200.0 & 1400.0 & $   882.0_{-   588.0}^{+   588.0}$ & 10.0 & 2002.4 \\
    CBI-M20h & [14] & 1502.0 & 1400.0 & 1600.0 & $  1353.0_{-   765.0}^{+   765.0}$ & 10.0 & 2002.4 \\
    CBI-M20h & [14] & 1702.0 & 1600.0 & 1800.0 & $   706.0_{-   706.0}^{+   706.0}$ & 10.0 & 2002.4 \\
    CBI-M20h & [14] & 1899.0 & 1800.0 & 2000.0 & $   118.0_{-   706.0}^{+   706.0}$ & 10.0 & 2002.4 \\
    COBE-DMR & [15] &    2.1 &    2.0 &    2.5 & $    72.2_{-    72.2}^{+   528.0}$ &  1.4 & 1996.0 \\
    COBE-DMR & [15] &    3.1 &    2.5 &    3.7 & $   784.0_{-   470.7}^{+   476.2}$ &  1.4 & 1996.0 \\
    COBE-DMR & [15] &    4.1 &    3.4 &    4.8 & $  1156.0_{-   437.8}^{+   444.0}$ &  1.4 & 1996.0 \\
    COBE-DMR & [15] &    5.6 &    4.7 &    6.6 & $   630.0_{-   287.8}^{+   294.1}$ &  1.4 & 1996.0 \\
    COBE-DMR & [15] &    8.0 &    6.8 &    9.3 & $   864.4_{-   224.3}^{+   224.6}$ &  1.4 & 1996.0 \\
    COBE-DMR & [15] &   10.9 &    9.7 &   12.2 & $   767.3_{-   229.1}^{+   231.3}$ &  1.4 & 1996.0 \\
    COBE-DMR & [15] &   14.4 &   12.8 &   15.7 & $   681.2_{-   244.4}^{+   249.0}$ &  1.4 & 1996.0 \\
    COBE-DMR & [15] &   19.4 &   16.6 &   22.1 & $  1089.0_{-   327.2}^{+   324.8}$ &  1.4 & 1996.0 \\
        DASI & [16] &  118.0 &  104.0 &  167.0 & $  3770.0_{-   820.0}^{+   820.0}$ &  8.0 & 2002.2 \\
        DASI & [16] &  203.0 &  173.0 &  255.0 & $  5280.0_{-   550.0}^{+   550.0}$ &  8.0 & 2002.2 \\
        DASI & [16] &  289.0 &  261.0 &  342.0 & $  3660.0_{-   340.0}^{+   340.0}$ &  8.0 & 2002.2 \\
        DASI & [16] &  377.0 &  342.0 &  418.0 & $  1650.0_{-   200.0}^{+   200.0}$ &  8.0 & 2002.2 \\
        DASI & [16] &  465.0 &  418.0 &  500.0 & $  1890.0_{-   220.0}^{+   220.0}$ &  8.0 & 2002.2 \\
        DASI & [16] &  553.0 &  506.0 &  594.0 & $  2840.0_{-   290.0}^{+   290.0}$ &  8.0 & 2002.2 \\
        DASI & [16] &  641.0 &  600.0 &  676.0 & $  1670.0_{-   270.0}^{+   270.0}$ &  8.0 & 2002.2 \\
        DASI & [16] &  725.0 &  676.0 &  757.0 & $  2010.0_{-   350.0}^{+   350.0}$ &  8.0 & 2002.2 \\
        DASI & [16] &  837.0 &  763.0 &  864.0 & $  2320.0_{-   450.0}^{+   450.0}$ &  8.0 & 2002.2 \\
        FIRS & [17] &   11.0 &    2.0 &   28.0 & $   864.4_{-   393.5}^{+   519.5}$ & -  &1994.7 \\
         IAB & [18] &  120.0 &   65.0 &  221.0 & $  8930.2_{-  6153.0}^{+  9647.4}$ & -  &1993.6 \\
      IACB94 & [19] &   33.0 &   17.0 &   59.0 & $ 12521.6_{-  9838.4}^{+ 18913.7}$ & 28.0 & 1998.3 \\
      IACB94 & [19] &   53.0 &   34.0 &   79.0 & $  2981.2_{-  1911.9}^{+  3710.1}$ & 28.0 & 1998.3 \\
      IACB96 & [20] &   39.0 &   15.0 &   77.0 & $  1156.0_{-   372.0}^{+   608.0}$ & 20.0 & 2001.1 \\
      IACB96 & [20] &   61.0 &   39.0 &   89.0 & $  1600.0_{-   444.0}^{+   609.0}$ & 20.0 & 2001.1 \\
      IACB96 & [20] &   81.0 &   61.0 &  108.0 & $  1681.0_{-   592.0}^{+   720.0}$ & 20.0 & 2001.1 \\
      IACB96 & [20] &   99.0 &   81.0 &  123.0 & $  2500.0_{-   819.0}^{+  1100.0}$ & 20.0 & 2001.1 \\
      IACB96 & [20] &  116.0 &  102.0 &  139.0 & $  2116.0_{-   747.0}^{+  1020.0}$ & 20.0 & 2001.1 \\
      IACB96 & [20] &  134.0 &  122.0 &  154.0 & $  3136.0_{-  1020.0}^{+  1353.0}$ & 20.0 & 2001.1 \\
       JBIAC & [21] &  109.0 &   90.0 &  128.0 & $  1849.0_{-   920.9}^{+  1309.9}$ & -  &1999.8 \\
     MAXIMA1 & [22] &   77.0 &   36.0 &  110.0 & $  1999.0_{-   506.0}^{+   675.0}$ &  8.0 & 2001.3 \\
     MAXIMA1 & [22] &  147.0 &  111.0 &  185.0 & $  2960.0_{-   554.0}^{+   682.0}$ &  8.0 & 2001.3 \\
     MAXIMA1 & [22] &  222.0 &  186.0 &  260.0 & $  6092.0_{-   901.0}^{+  1052.0}$ &  8.0 & 2001.3 \\
     MAXIMA1 & [22] &  294.0 &  261.0 &  335.0 & $  3830.0_{-   577.0}^{+   670.0}$ &  8.0 & 2001.3 \\
     MAXIMA1 & [22] &  381.0 &  336.0 &  410.0 & $  2270.0_{-   471.0}^{+   569.0}$ &  8.0 & 2001.3 \\
     MAXIMA1 & [22] &  449.0 &  411.0 &  485.0 & $  1468.0_{-   325.0}^{+   387.0}$ &  8.0 & 2001.3 \\
     MAXIMA1 & [22] &  523.0 &  486.0 &  560.0 & $  1935.0_{-   408.0}^{+   475.0}$ &  8.0 & 2001.3 \\
     MAXIMA1 & [22] &  597.0 &  561.0 &  635.0 & $  1811.0_{-   441.0}^{+   511.0}$ &  8.0 & 2001.3 \\
     MAXIMA1 & [22] &  671.0 &  636.0 &  710.0 & $  2100.0_{-   546.0}^{+   629.0}$ &  8.0 & 2001.3 \\
     MAXIMA1 & [22] &  746.0 &  711.0 &  785.0 & $  2189.0_{-   680.0}^{+   777.0}$ &  8.0 & 2001.3 \\
     MAXIMA1 & [22] &  856.0 &  786.0 &  935.0 & $  3104.0_{-   738.0}^{+   805.0}$ &  8.0 & 2001.3 \\
     MAXIMA1 & [22] & 1004.0 &  936.0 & 1085.0 & $  1084.0_{-  1085.0}^{+  1219.0}$ &  8.0 & 2001.3 \\
     MAXIMA1 & [22] & 1147.0 & 1086.0 & 1235.0 & $   223.0_{-  2025.0}^{+  2791.0}$ &  8.0 & 2001.3 \\
        MSAM & [23] &   84.0 &   39.0 &  130.0 & $  1225.0_{-   649.0}^{+  1275.0}$ & 10.0 & 2000.2 \\
        MSAM & [23] &  201.0 &  131.0 &  283.0 & $  2401.0_{-   720.0}^{+  1080.0}$ & 10.0 & 2000.2 \\
        MSAM & [23] &  407.0 &  284.0 &  453.0 & $  2209.0_{-   528.0}^{+   707.0}$ & 10.0 & 2000.2 \\
        OVRO & [24] &  589.0 &  361.0 &  756.0 & $  3481.0_{-   735.7}^{+  1077.3}$ & -  &2000.2 \\
     PyI-III & [25] &   87.0 &   49.0 &  105.0 & $  3600.0_{-   575.0}^{+  1161.0}$ & 40.0 & 1997.0 \\
     PyI-III & [25] &  170.0 &  120.0 &  239.0 & $  4356.0_{-  1107.0}^{+  1527.0}$ & 40.0 & 1997.0 \\
         PyV & [26] &   44.0 &   29.0 &   59.0 & $   484.0_{-   195.0}^{+   192.0}$ & 30.0 & 2003.1 \\
         PyV & [26] &   75.0 &   60.0 &   90.0 & $   576.0_{-   287.0}^{+   324.0}$ & 30.0 & 2003.1 \\
         PyV & [26] &  106.0 &   91.0 &  121.0 & $  1156.0_{-   531.0}^{+   525.0}$ & 30.0 & 2003.1 \\
         PyV & [26] &  137.0 &  122.0 &  152.0 & $  2500.0_{-  1056.0}^{+   981.0}$ & 30.0 & 2003.1 \\
         PyV & [26] &  168.0 &  153.0 &  183.0 & $  3721.0_{-  1785.0}^{+  1755.0}$ & 30.0 & 2003.1 \\
         PyV & [26] &  199.0 &  184.0 &  214.0 & $  5929.0_{-  3528.0}^{+  3480.0}$ & 30.0 & 2003.1 \\
         PyV & [26] &  230.0 &  215.0 &  245.0 & $     0.0_{-     0.0}^{+  7569.0}$ & 30.0 & 2003.1 \\
         PyV & [26] &  261.0 &  246.0 &  276.0 & $  4761.0_{-  4761.0}^{+ 14839.0}$ & 30.0 & 2003.1 \\
        QMAP & [27] &   80.0 &   39.0 &  121.0 & $  2401.0_{-   679.0}^{+   668.0}$ & 16.0 & 2002.4 \\
        QMAP & [27] &  126.0 &   72.0 &  180.0 & $  3069.0_{-   559.0}^{+   615.0}$ & 16.0 & 2002.4 \\
        QMAP & [27] &  111.0 &   47.0 &  175.0 & $  3819.0_{-   871.0}^{+   832.0}$ & 16.0 & 2002.4 \\
          SK & [27] &   87.0 &   58.0 &  126.0 & $  2520.0_{-   495.0}^{+   902.0}$ & 20.0 & 2002.4 \\
          SK & [27] &  166.0 &  123.0 &  196.0 & $  4970.0_{-   836.0}^{+  1080.0}$ & 20.0 & 2002.4 \\
          SK & [27] &  237.0 &  196.0 &  266.0 & $  7535.0_{-  1372.0}^{+  1914.0}$ & 20.0 & 2002.4 \\
          SK & [27] &  286.0 &  248.0 &  310.0 & $  7726.0_{-  1720.0}^{+  2354.0}$ & 20.0 & 2002.4 \\
          SK & [27] &  349.0 &  308.0 &  393.0 & $  4956.0_{-  3275.0}^{+  3180.0}$ & 20.0 & 2002.4 \\
    Tenerife & [28] &   20.0 &   12.0 &   30.0 & $   900.0_{-   553.7}^{+  1112.7}$ & -  &2000.0 \\
      TOCO97 & [27] &   63.0 &   45.0 &   81.0 & $  1232.0_{-   408.0}^{+   820.0}$ & 20.0 & 2002.4 \\
      TOCO97 & [27] &   86.0 &   64.0 &  102.0 & $  1846.0_{-   499.0}^{+   644.0}$ & 20.0 & 2002.4 \\
      TOCO97 & [27] &  114.0 &   90.0 &  134.0 & $  4529.0_{-   747.0}^{+   888.0}$ & 20.0 & 2002.4 \\
      TOCO97 & [27] &  158.0 &  135.0 &  180.0 & $  7465.0_{-  1177.0}^{+  1296.0}$ & 20.0 & 2002.4 \\
      TOCO97 & [27] &  199.0 &  170.0 &  237.0 & $  6872.0_{-  1202.0}^{+  1318.0}$ & 20.0 & 2002.4 \\
      TOCO98 & [27] &  128.0 &   95.0 &  154.0 & $  2884.0_{-  1485.0}^{+  2272.0}$ & 16.0 & 2002.4 \\
      TOCO98 & [27] &  152.0 &  114.0 &  178.0 & $  6497.0_{-  1625.0}^{+  1858.0}$ & 16.0 & 2002.4 \\
      TOCO98 & [27] &  226.0 &  170.0 &  263.0 & $  6659.0_{-  1094.0}^{+  1174.0}$ & 16.0 & 2002.4 \\
      TOCO98 & [27] &  306.0 &  247.0 &  350.0 & $  4733.0_{-  1369.0}^{+  1445.0}$ & 16.0 & 2002.4 \\
      TOCO98 & [27] &  409.0 &  344.0 &  451.0 & $   545.0_{-  2043.0}^{+  2043.0}$ & 16.0 & 2002.4 \\
       VIPER & [29] &  108.0 &   30.0 &  228.0 & $  3721.0_{-  2200.0}^{+  4743.0}$ & 16.0 & 2000.3 \\
       VIPER & [29] &  173.0 &   73.0 &  288.0 & $  5929.0_{-  2680.0}^{+  4680.0}$ & 16.0 & 2000.3 \\
       VIPER & [29] &  237.0 &  126.0 &  336.0 & $  4225.0_{-  1921.0}^{+  3696.0}$ & 16.0 & 2000.3 \\
       VIPER & [29] &  263.0 &  150.0 &  448.0 & $  6241.0_{-  2016.0}^{+  3168.0}$ & 16.0 & 2000.3 \\
       VIPER & [29] &  422.0 &  291.0 &  604.0 & $   784.0_{-   615.0}^{+  1065.0}$ & 16.0 & 2000.3 \\
       VIPER & [29] &  589.0 &  448.0 &  796.0 & $  4225.0_{-  2625.0}^{+  3875.0}$ & 16.0 & 2000.3 \\
         VSA & [30] &  160.0 &  100.0 &  190.0 & $  3864.0_{-  1141.0}^{+  1587.0}$ &  7.0 & 2002.9 \\
         VSA & [30] &  220.0 &  190.0 &  250.0 & $  5893.0_{-  1339.0}^{+  1637.0}$ &  7.0 & 2002.9 \\
         VSA & [30] &  289.0 &  250.0 &  310.0 & $  5390.0_{-  1091.0}^{+  1289.0}$ &  7.0 & 2002.9 \\
         VSA & [30] &  349.0 &  310.0 &  370.0 & $  2603.0_{-   545.0}^{+   595.0}$ &  7.0 & 2002.9 \\
         VSA & [30] &  416.0 &  370.0 &  450.0 & $  1749.0_{-   347.0}^{+   347.0}$ &  7.0 & 2002.9 \\
         VSA & [30] &  479.0 &  450.0 &  500.0 & $  1638.0_{-   496.0}^{+   644.0}$ &  7.0 & 2002.9 \\
         VSA & [30] &  537.0 &  500.0 &  580.0 & $  2866.0_{-   545.0}^{+   595.0}$ &  7.0 & 2002.9 \\
         VSA & [30] &  605.0 &  580.0 &  640.0 & $  1460.0_{-   545.0}^{+   644.0}$ &  7.0 & 2002.9 \\
         VSA & [30] &  670.0 &  640.0 &  700.0 & $  2237.0_{-   595.0}^{+   694.0}$ &  7.0 & 2002.9 \\
         VSA & [30] &  726.0 &  700.0 &  750.0 & $  1922.0_{-   744.0}^{+   793.0}$ &  7.0 & 2002.9 \\
         VSA & [30] &  795.0 &  750.0 &  850.0 & $  3587.0_{-   644.0}^{+   644.0}$ &  7.0 & 2002.9 \\
         VSA & [30] &  888.0 &  850.0 &  950.0 & $  1471.0_{-   545.0}^{+   644.0}$ &  7.0 & 2002.9 \\
         VSA & [30] & 1002.0 &  950.0 & 1050.0 & $     0.0_{-     0.0}^{+  1091.0}$ &  7.0 & 2002.9 \\
         VSA & [30] & 1119.0 & 1050.0 & 1200.0 & $  1125.0_{-   644.0}^{+   694.0}$ &  7.0 & 2002.9 \\
         VSA & [30] & 1271.0 & 1200.0 & 1350.0 & $     0.0_{-     0.0}^{+  1431.0}$ &  7.0 & 2002.9 \\
         VSA & [30] & 1419.0 & 1350.0 & 1700.0 & $  1311.0_{-  1289.0}^{+  1538.0}$ &  7.0 & 2002.9 \\
        WMAP & [31] &    2.0 &    2.0 &    2.0 & $   123.4_{-   762.6}^{+   762.6}$ &  1.0 & 2003.1 \\
        WMAP & [31] &    3.0 &    3.0 &    3.0 & $   611.8_{-   608.2}^{+   608.2}$ &  1.0 & 2003.1 \\
        WMAP & [31] &    4.0 &    4.0 &    4.0 & $   756.6_{-   504.0}^{+   504.0}$ &  1.0 & 2003.1 \\
        WMAP & [31] &    5.0 &    5.0 &    5.0 & $  1256.7_{-   432.2}^{+   432.2}$ &  1.0 & 2003.1 \\
        WMAP & [31] &    6.0 &    6.0 &    6.0 & $   696.5_{-   380.2}^{+   380.2}$ &  1.0 & 2003.1 \\
        WMAP & [31] &    7.0 &    7.0 &    7.0 & $   829.8_{-   342.6}^{+   342.6}$ &  1.0 & 2003.1 \\
        WMAP & [31] &    8.0 &    8.0 &    8.0 & $   627.9_{-   314.4}^{+   314.4}$ &  1.0 & 2003.1 \\
        WMAP & [31] &    9.0 &    9.0 &    9.0 & $   815.2_{-   292.5}^{+   292.5}$ &  1.0 & 2003.1 \\
        WMAP & [31] &   10.0 &   10.0 &   10.0 & $   617.8_{-   275.5}^{+   275.5}$ &  1.0 & 2003.1 \\
        WMAP & [31] &   11.5 &   11.0 &   12.0 & $   995.8_{-   182.1}^{+   182.1}$ &  1.0 & 2003.1 \\
        WMAP & [31] &   13.5 &   13.0 &   14.0 & $   813.4_{-   170.7}^{+   170.7}$ &  1.0 & 2003.1 \\
        WMAP & [31] &   15.5 &   15.0 &   16.0 & $   748.1_{-   163.2}^{+   163.2}$ &  1.0 & 2003.1 \\
        WMAP & [31] &   17.5 &   17.0 &   18.0 & $   890.6_{-   158.3}^{+   158.3}$ &  1.0 & 2003.1 \\
        WMAP & [31] &   19.5 &   19.0 &   20.0 & $   908.7_{-   155.4}^{+   155.4}$ &  1.0 & 2003.1 \\
        WMAP & [31] &   23.0 &   21.0 &   25.0 & $   722.2_{-    96.7}^{+    96.7}$ &  1.0 & 2003.1 \\
        WMAP & [31] &   28.0 &   26.0 &   30.0 & $  1055.9_{-    96.4}^{+    96.4}$ &  1.0 & 2003.1 \\
        WMAP & [31] &   33.0 &   31.0 &   35.0 & $  1171.4_{-    97.4}^{+    97.4}$ &  1.0 & 2003.1 \\
        WMAP & [31] &   38.0 &   36.0 &   40.0 & $  1422.8_{-    98.9}^{+    98.9}$ &  1.0 & 2003.1 \\
        WMAP & [31] &   43.0 &   41.0 &   45.0 & $  1280.6_{-   100.9}^{+   100.9}$ &  1.0 & 2003.1 \\
        WMAP & [31] &   48.0 &   46.0 &   50.0 & $  1323.7_{-   103.3}^{+   103.3}$ &  1.0 & 2003.1 \\
        WMAP & [31] &   53.0 &   51.0 &   55.0 & $  1313.9_{-   106.1}^{+   106.1}$ &  1.0 & 2003.1 \\
        WMAP & [31] &   58.0 &   56.0 &   60.0 & $  1612.2_{-   109.3}^{+   109.3}$ &  1.0 & 2003.1 \\
        WMAP & [31] &   63.0 &   61.0 &   65.0 & $  1626.0_{-   112.9}^{+   112.9}$ &  1.0 & 2003.1 \\
        WMAP & [31] &   68.0 &   66.0 &   70.0 & $  1828.4_{-   116.7}^{+   116.7}$ &  1.0 & 2003.1 \\
        WMAP & [31] &   73.0 &   71.0 &   75.0 & $  1968.6_{-   120.8}^{+   120.8}$ &  1.0 & 2003.1 \\
        WMAP & [31] &   78.0 &   76.0 &   80.0 & $  1888.3_{-   125.1}^{+   125.1}$ &  1.0 & 2003.1 \\
        WMAP & [31] &   83.0 &   81.0 &   85.0 & $  2223.7_{-   129.7}^{+   129.7}$ &  1.0 & 2003.1 \\
        WMAP & [31] &   88.0 &   86.0 &   90.0 & $  2299.0_{-   134.3}^{+   134.3}$ &  1.0 & 2003.1 \\
        WMAP & [31] &   93.0 &   91.0 &   95.0 & $  2409.3_{-   139.2}^{+   139.2}$ &  1.0 & 2003.1 \\
        WMAP & [31] &   98.0 &   96.0 &  100.0 & $  2504.5_{-   144.3}^{+   144.3}$ &  1.0 & 2003.1 \\
        WMAP & [31] &  105.5 &  101.0 &  110.0 & $  2863.1_{-    97.5}^{+    97.5}$ &  1.0 & 2003.1 \\
        WMAP & [31] &  115.5 &  111.0 &  120.0 & $  3165.6_{-   104.1}^{+   104.1}$ &  1.0 & 2003.1 \\
        WMAP & [31] &  125.5 &  121.0 &  130.0 & $  3438.4_{-   110.9}^{+   110.9}$ &  1.0 & 2003.1 \\
        WMAP & [31] &  135.5 &  131.0 &  140.0 & $  3766.5_{-   117.4}^{+   117.4}$ &  1.0 & 2003.1 \\
        WMAP & [31] &  145.5 &  141.0 &  150.0 & $  4105.7_{-   123.6}^{+   123.6}$ &  1.0 & 2003.1 \\
        WMAP & [31] &  155.5 &  151.0 &  160.0 & $  4531.5_{-   129.1}^{+   129.1}$ &  1.0 & 2003.1 \\
        WMAP & [31] &  165.5 &  161.0 &  170.0 & $  4539.6_{-   133.9}^{+   133.9}$ &  1.0 & 2003.1 \\
        WMAP & [31] &  175.5 &  171.0 &  180.0 & $  5030.4_{-   137.8}^{+   137.8}$ &  1.0 & 2003.1 \\
        WMAP & [31] &  185.5 &  181.0 &  190.0 & $  5090.9_{-   140.7}^{+   140.7}$ &  1.0 & 2003.1 \\
        WMAP & [31] &  195.5 &  191.0 &  200.0 & $  5529.6_{-   142.7}^{+   142.7}$ &  1.0 & 2003.1 \\
        WMAP & [31] &  210.5 &  201.0 &  220.0 & $  5366.5_{-   101.9}^{+   101.9}$ &  1.0 & 2003.1 \\
        WMAP & [31] &  230.5 &  221.0 &  240.0 & $  5654.5_{-    99.2}^{+    99.2}$ &  1.0 & 2003.1 \\
        WMAP & [31] &  250.5 &  241.0 &  260.0 & $  5271.3_{-    93.8}^{+    93.8}$ &  1.0 & 2003.1 \\
        WMAP & [31] &  270.5 &  261.0 &  280.0 & $  4877.8_{-    86.3}^{+    86.3}$ &  1.0 & 2003.1 \\
        WMAP & [31] &  290.5 &  281.0 &  300.0 & $  4344.4_{-    78.1}^{+    78.1}$ &  1.0 & 2003.1 \\
        WMAP & [31] &  310.5 &  301.0 &  320.0 & $  3536.5_{-    70.4}^{+    70.4}$ &  1.0 & 2003.1 \\
        WMAP & [31] &  330.5 &  321.0 &  340.0 & $  2936.5_{-    64.5}^{+    64.5}$ &  1.0 & 2003.1 \\
        WMAP & [31] &  350.5 &  341.0 &  360.0 & $  2266.5_{-    61.2}^{+    61.2}$ &  1.0 & 2003.1 \\
        WMAP & [31] &  370.5 &  361.0 &  380.0 & $  2007.2_{-    61.2}^{+    61.2}$ &  1.0 & 2003.1 \\
        WMAP & [31] &  390.5 &  381.0 &  400.0 & $  1592.7_{-    64.8}^{+    64.8}$ &  1.0 & 2003.1 \\
        WMAP & [31] &  410.5 &  401.0 &  420.0 & $  1769.5_{-    72.0}^{+    72.0}$ &  1.0 & 2003.1 \\
        WMAP & [31] &  430.5 &  421.0 &  440.0 & $  1801.2_{-    82.4}^{+    82.4}$ &  1.0 & 2003.1 \\
        WMAP & [31] &  450.5 &  441.0 &  460.0 & $  1832.9_{-    94.9}^{+    94.9}$ &  1.0 & 2003.1 \\
        WMAP & [31] &  470.5 &  461.0 &  480.0 & $  2120.1_{-   109.5}^{+   109.5}$ &  1.0 & 2003.1 \\
        WMAP & [31] &  490.5 &  481.0 &  500.0 & $  2246.0_{-   127.0}^{+   127.0}$ &  1.0 & 2003.1 \\
        WMAP & [31] &  513.0 &  501.0 &  525.0 & $  2237.3_{-   133.3}^{+   133.3}$ &  1.0 & 2003.1 \\
        WMAP & [31] &  538.0 &  526.0 &  550.0 & $  2729.9_{-   158.1}^{+   158.1}$ &  1.0 & 2003.1 \\
        WMAP & [31] &  563.0 &  551.0 &  575.0 & $  2274.0_{-   186.8}^{+   186.8}$ &  1.0 & 2003.1 \\
        WMAP & [31] &  588.0 &  576.0 &  600.0 & $  2330.5_{-   220.7}^{+   220.7}$ &  1.0 & 2003.1 \\
        WMAP & [31] &  613.0 &  601.0 &  625.0 & $  1802.1_{-   261.4}^{+   261.4}$ &  1.0 & 2003.1 \\
        WMAP & [31] &  638.0 &  626.0 &  650.0 & $  1871.8_{-   310.8}^{+   310.8}$ &  1.0 & 2003.1 \\
        WMAP & [31] &  663.0 &  651.0 &  675.0 & $  2301.3_{-   371.0}^{+   371.0}$ &  1.0 & 2003.1 \\
        WMAP & [31] &  688.0 &  676.0 &  700.0 & $  1560.8_{-   444.1}^{+   444.1}$ &  1.0 & 2003.1 \\
        WMAP & [31] &  713.0 &  701.0 &  725.0 & $   924.1_{-   532.2}^{+   532.2}$ &  1.0 & 2003.1 \\
        WMAP & [31] &  738.0 &  726.0 &  750.0 & $  2280.3_{-   637.5}^{+   637.5}$ &  1.0 & 2003.1 \\
        WMAP & [31] &  763.0 &  751.0 &  775.0 & $  2080.8_{-   762.3}^{+   762.3}$ &  1.0 & 2003.1 \\
        WMAP & [31] &  788.0 &  776.0 &  800.0 & $  1510.9_{-   909.5}^{+   909.5}$ &  1.0 & 2003.1 \\
        WMAP & [31] &  813.0 &  801.0 &  825.0 & $  1250.7_{-  1082.1}^{+  1082.1}$ &  1.0 & 2003.1 \\
        WMAP & [31] &  838.0 &  826.0 &  850.0 & $  3539.0_{-  1283.9}^{+  1283.9}$ &  1.0 & 2003.1 \\
        WMAP & [31] &  875.5 &  851.0 &  900.0 & $  3380.4_{-  1159.4}^{+  1159.4}$ &  1.0 & 2003.1 \\
\enddata 
\tablerefs{[1] \cite{kuo02}, [2] \cite{tanaka96}, \cite{lineweaver98}, [3] \cite{benoit02}, [4] \cite{gunderson95}, [5] \cite{debernardis94}, [6] \cite{masi96}, [7] \cite{tucker97}, [8] \cite{mauskopf00}, [9] \cite{ruhl02}, [10] \cite{scott96}, [11] \cite{baker99}, [12] \cite{padin01}, [13] \cite{mason02}, [14] \cite{pearson02}, [15] \cite{tegmark97}, [16] \cite{halverson02}, [17] \cite{ganga94a}, [18] \cite{piccirillo93}, [19] \cite{femenia98}, [20] \cite{romeo01}, [21] \cite{dicker99}, [22] \cite{lee01}, [23] \cite{wilson00}, [24] \cite{leitch00}, [25] \cite{platt97}, [26] \cite{coble03}, [27] \cite{miller02}, [28] \cite{gutierrez00}, [29] \cite{peterson00}, [30] \cite{grainge02}, [31] \cite{hinshaw03}.}
\tablenotetext{a}{The 1$\sigma$ calibration uncertainty in temperature, $\sigma_u$, is given as a percentage and allows the data points taken at the same time using the same instrument to shift upwards or downwards together.  For observations that result in a single data point, $\sigma_u$ is not given since either it is not quoted in the literature, or it has been treated by adding it in quadrature to the statistical error bars.}
\end{deluxetable}

\clearpage
\begin{deluxetable}{lccccccccccccc}
\rotate
\tabletypesize{\tiny}
\tablecolumns{10} 
\tablewidth{0pc}
\setlength{\tabcolsep}{0.16in}
\tablecaption{\label{obstech} Details of the CMB observational techniques.} 
\tablehead{ 
\colhead{Experiment} & \colhead{Ref.}   & \colhead{$\nu_{\rm main} \pm \Delta \nu_{\rm main}$} & \colhead{$\nu$ Range}  & \colhead{$\theta_{\rm beam}$ FWHM}  & \colhead{Point. Uncert.$^a$} & 
 \colhead{Calibration}  & \colhead{Instrument}  & \colhead{Platform} & \colhead{$|$Gal. Lat.$|$} \\
\colhead{} & \colhead{}   & \colhead{(GHz)} & \colhead{(GHz)}  & \colhead{(arcmin)}  & \colhead{(arcmin)} & 
 \colhead{Source}  & \colhead{Type}  & \colhead{}  & \colhead{Range (deg)}}
\startdata 
       ACBAR &  [1] & 150.0 $\pm$ 15.0 & 150.0 - 280.0 &   4.5 &  0.30 & Venus \& Mars &  Bolometer &     Ground &  36.7 -  57.0 \\
    ACME-MAX &  [2] & 180.0 $\pm$  7.0 & 105.0 - 420.0 &  30.0 &  1.00 &       Jupiter &  Bolometer &    Balloon &  40.8 -  76.6 \\
   ACME-SP91 &  [3] &  27.7 $\pm$  1.2 &  27.7 -  27.7 &  96.0 &  5.00 &      Taurus A &       HEMT &     Ground &  45.0 -  55.0 \\
   ACME-SP94 &  [4] &  35.0 $\pm$  1.2 &  27.7 -  41.5 &  83.0 &  7.20 &      H/C Load &       HEMT &     Ground &  40.0 -  55.0 \\
    ARCHEOPS &  [5] & 190.0 $\pm$ 31.0 & 143.0 - 545.0 &   8.0 &  1.50 &        Dipole &  Bolometer &    Balloon &  30.0 -  90.0 \\
       ARGO1 &  [6] & 150.0 $\pm$ 15.0 & 150.0 - 600.0 &  52.0 &  2.00 &      H/C Load &  Bolometer &    Balloon &  22.0 -  35.0 \\
       ARGO2 &  [7] & 150.0 $\pm$ 15.0 & 150.0 - 600.0 &  52.0 &  2.00 &      H/C Load &  Bolometer &    Balloon &   0.0 -   7.8 \\
         BAM &  [8] & 147.0 $\pm$ 20.0 & 111.0 - 255.0 &  42.0 &  3.00 &       Jupiter &    Bol/Int &    Balloon &  12.8 -  32.8 \\
      BOOM97 &  [9] & 153.0 $\pm$ 21.0 &  96.0 - 153.0 &  18.0 &  1.00 &       Jupiter &  Bolometer &    Balloon &  11.0 -  83.0 \\
      BOOM98 & [10] & 150.0 $\pm$ 11.0 &  90.0 - 410.0 &  11.1 &  2.50 &        Dipole &  Bolometer &    Balloon &  18.0 -  45.0 \\
        CAT1 & [11] &  16.5 $\pm$  0.2 &  13.5 -  16.5 & 117.6 & - &         Cas A &   HEMT/Int &     Ground &  29.5 -  38.0 \\
        CAT2 & [12] &  16.5 $\pm$  0.2 &  13.5 -  16.5 & 117.6 & - &         Cas A &   HEMT/Int &     Ground &  33.9 -  40.4 \\
      CBI-1a & [13] &  31.0 $\pm$  0.5 &  26.5 -  35.5 &   3.0 &  0.05 &      Taurus A &   HEMT/Int &     Ground &  23.3 -  25.0 \\
      CBI-1b & [13] &  31.0 $\pm$  0.5 &  26.5 -  35.5 &   3.0 &  0.05 &      Taurus A &   HEMT/Int &     Ground &  47.8 -  49.1 \\
     CBI-D8h & [14] &  31.0 $\pm$  0.5 &  26.5 -  35.5 &   3.0 &  0.03 &       Jupiter &   HEMT/Int &     Ground &  23.3 -  25.0 \\
    CBI-D14h & [14] &  31.0 $\pm$  0.5 &  26.5 -  35.5 &   3.0 &  0.03 &       Jupiter &   HEMT/Int &     Ground &  53.4 -  54.8 \\
    CBI-D20h & [14] &  31.0 $\pm$  0.5 &  26.5 -  35.5 &   3.0 &  0.03 &       Jupiter &   HEMT/Int &     Ground &  27.6 -  29.3 \\
     CBI-M2h & [15] &  31.0 $\pm$  0.5 &  26.5 -  35.5 &   3.0 &  0.03 &       Jupiter &   HEMT/Int &     Ground &  53.0 -  55.0 \\
    CBI-M14h & [15] &  31.0 $\pm$  0.5 &  26.5 -  35.5 &   3.0 &  0.03 &       Jupiter &   HEMT/Int &     Ground &  48.0 -  50.0 \\
    CBI-M20h & [15] &  31.0 $\pm$  0.5 &  26.5 -  35.5 &   3.0 &  0.03 &       Jupiter &   HEMT/Int &     Ground &  26.0 -  28.0 \\
    COBE-DMR & [16] &  53.0 $\pm$  0.1 &  31.5 -  90.0 & 420.0 &  3.00 &      H/C Load &       HEMT &  Satellite &  20.0 -  90.0 \\
        DASI & [17] &  31.0 $\pm$  0.5 &  26.5 -  35.5 &  20.0 &  2.00 &  Gal. Sources &   HEMT/Int &     Ground &  26.9 -  67.3 \\
        FIRS & [18] & 167.0 $\pm$ 19.0 & 167.0 - 682.0 & 228.0 & 60.00 &        Dipole &  Bolometer &    Balloon &   0.0 -  80.0 \\
         IAB & [19] & 136.0 $\pm$  1.5 & 136.0 - 136.0 &  50.0 &  2.00 &      H/C Load &  Bolometer &     Ground &  22.2 -  32.1 \\
      IACB94 & [20] & 116.0 $\pm$  1.5 &  90.9 - 272.7 & 121.8 &  5.40 &          Moon &  Bolometer &     Ground &   0.0 -  49.2 \\
      IACB96 & [21] & 142.9 $\pm$  1.5 &  96.8 - 272.7 &  81.0 &  5.40 &          Moon &  Bolometer &     Ground &   0.0 -  87.8 \\
       JBIAC & [22] &  33.0 $\pm$  1.5 &  33.0 -  33.0 & 120.0 & - &          Moon &   HEMT/Int &     Ground &   0.0 -  76.0 \\
     MAXIMA1 & [23] & 150.0 $\pm$ 35.0 & 150.0 - 410.0 &  10.0 &  0.95 &        Dipole &  Bolometer &    Balloon &  44.0 -  54.3 \\
        MSAM & [24] & 170.0 $\pm$ 22.5 & 170.0 - 680.0 &  30.0 &  2.50 &       Jupiter &  Bolometer &    Balloon &  25.0 -  36.0 \\
        OVRO & [25] &  31.7 $\pm$  3.0 &  14.5 -  31.7 &   7.4 &  2.00 &       Jupiter &       HEMT &     Ground &  25.1 -  29.3 \\
     PyI-III & [26] &  90.0 $\pm$ 18.0 &  90.0 -  90.0 &  45.0 &  6.00 &      H/C Load &  Bolometer &     Ground &  60.0 -  70.0 \\
         PyV & [27] &  40.3 $\pm$  2.8 &  40.3 -  40.3 &  60.0 &  9.00 &      H/C Load &       HEMT &     Ground &  30.0 -  70.0 \\
        QMAP & [28] &  37.0 $\pm$  3.1 &  31.0 -  42.0 &  48.0 &  3.60 &         Cas A &       HEMT &    Balloon &   8.0 -  46.0 \\
          SK & [29] &  42.0 $\pm$  3.5 &  31.0 -  42.0 &  28.0 &  1.80 &         Cas A &       HEMT &     Ground &  19.0 -  35.0 \\
    Tenerife & [30] &  15.0 $\pm$  0.8 &  10.0 -  15.0 & 300.0 & - &   Combination &       HEMT &     Ground &  40.0 -  90.0 \\
      TOCO97 & [31] &  37.0 $\pm$  3.0 &  31.0 - 144.0 &  48.0 &  0.45 &       Jupiter &   HEMT/SIS &     Ground &   0.0 -  55.0 \\
      TOCO98 & [31] & 144.0 $\pm$  1.7 &  31.0 - 144.0 &  12.0 &  0.45 &       Jupiter &   HEMT/SIS &     Ground &   0.0 -  55.0 \\
       VIPER & [32] &  40.0 $\pm$  3.0 &  40.0 -  40.0 &  15.6 &  4.00 &      H/C Load &       HEMT &     Ground &  50.0 -  60.0 \\
         VSA & [33] &  34.0 $\pm$  0.8 &  34.0 -  34.0 & 120.0 &  5.00 &       Jupiter &   HEMT/Int &     Ground &  31.5 -  54.3 \\
        WMAP & [34] &  61.0 $\pm$  7.0 &  23.0 -  93.0 &  21.0 &  5.00 &        Dipole &       HEMT &  Satellite &  20.0 -  90.0 \\
\enddata 
\tablerefs{[1] \cite{kuo02}, [2] \cite{alsop92}, \cite{lim96}, \cite{tanaka96}, [3] \cite{gunderson95}, [4] \cite{benoit02} [5] \cite{ganga94b}, \cite{gunderson95}, [6] \cite{debernardis93}, \cite{debernardis94}, [7] \cite{debernardis93}, \cite{debernardis94}, \cite{masi95}, \cite{masi96}, [8] \cite{tucker97}, [9] \cite{mauskopf00}, \cite{piacentini02}, [10] \cite{crill02}, \cite{netterfield02}, \cite{ruhl02}, [11] \cite{scott96}, [12] \cite{baker99}, [13] \cite{padin01}, \cite{padin02}, [14] \cite{mason02}, [15] \cite{mason02}, \cite{pearson02}, [16] \cite{kogut92}, \cite{kogut96}, \cite{tegmark97}, [17] \cite{halverson02}, \cite{leitch02}, [18] \cite{page90}, \cite{meyer91}, \cite{ganga94a}, [19] \cite{piccirillo93}, [20] \cite{femenia98}, [21] \cite{femenia98}, \cite{romeo01}, [22] \cite{dicker99}, \cite{melhuish99}, [23] \cite{lee01}, \cite{hanany00}, [24] \cite{fixsen96}, \cite{wilson00}, [25] \cite{leitch00}, [26] \cite{dragovan94}, \cite{ruhl95}, \cite{platt97}, [27] \cite{coble99}, \cite{coble03}, [28] \cite{deoliveira98}, \cite{devlin98}, \cite{herbig98}, \cite{miller02}, [29] \cite{netterfield97}, \cite{miller02}, [30] \cite{davies96}, \cite{gutierrez00}, [31] \cite{miller02}, [32] \cite{peterson00}, [33] \cite{grainge02}, \cite{scott02}, \cite{taylor02}, \cite{watson02}, [34] \cite{bennett03}, \cite{hinshaw03}.}
\vskip -0.2cm
\tablenotetext{a}{Where pointing uncertainties are not given, they are not quoted in the literature.}
\vskip -0.2cm
\tablenotetext{b}{The VSA results quoted in the literature are the combined detections from a number of separate fields observed at various galactic longitudes and latitudes.  Information is not given in the literature to enable the contributions from the different fields to be separated. Therefore, the VSA galactic longitude range is omitted.  However, since the fields are not very dispersed in galactic latitude, this range is listed.}
\vskip -0.2cm
\end{deluxetable}
\clearpage
\begin{deluxetable}{lccccccccccccccc} 
\rotate
\tablecolumns{12} 
\tablewidth{0pc} 
\setlength{\tabcolsep}{0.04in}
\tablecaption{\label{resres} Results of the residual analyses.}
\tablehead{ 
\colhead{} & \colhead{$\ell_{\rm eff}$} & \colhead{$\begin{array}{c}|b| \\ {\rm range}^a\end{array}$} & \colhead{$\begin{array}{c}\rm central \\|b|^b \end{array}$} & \colhead{$\nu_{\rm main}$} & \colhead{$\frac{(\nu_{\rm max}-\nu_{\rm min})}{\nu_{\rm main}}$} & \colhead{$\begin{array}{c}\rm cal.\\ \rm source\end{array}$} & \colhead{$\begin{array}{c}\rm inst.\\ \rm type\end{array}$} & \colhead{$\theta_{\rm beam}\ell_{\rm eff}$} & \colhead{$\begin{array}{c}\rm point.\\ \rm uncert\end{array}$} &\colhead{$\begin{array}{c}\rm inst. \\ \rm platform\end{array}$}& \colhead{$\frac{\Delta\ell}{\ell_{eff}}$}}
\startdata
Figure & 3 & 4 & 5 & 6 & 7 & 8 & 9 & 10 & 11 & - & - \\
$\chi^2$ per dof & 1.06 & 0.74 & 1.06 & 1.06 & 1.06 & 1.06 & 1.06 & 1.06 & 1.07 & 1.06 & 1.05 \\
significance$^c$ & $\sim 1/2\sigma$ & $>3\sigma$ &$<\frac{1}{2}\sigma$ &$<\frac{1}{2}\sigma$ &$<\frac{1}{2}\sigma$ &$ <1\sigma$&$<\frac{1}{2}\sigma$&$<\frac{1}{2}\sigma$&$<\frac{1}{2}\sigma$&$<\frac{1}{2}\sigma$&$<2\sigma$ \\
\enddata 
\tablenotetext{a}{The $|b|$ range analysis involves a 2 dimensional fit (see \S5).}
\tablenotetext{b}{The central $|b|$ analysis involves a 1 dimensional fit (see \S5).}
\tablenotetext{c}{The significance of the deviation of the zero-line from the best-fit linear model.  As discussed in \S2.2, correlations other than the correlated beam and calibration uncertainties of individual experiments are not considered in our analysis. Ignoring such correlations may result in the significance of the deviation being smaller than expected.}
\end{deluxetable}

\end{document}